\documentclass[
amsmath,amssymb, two column
 ]{revtex4}
\usepackage{graphicx}
\usepackage{dcolumn}
\usepackage{bm}
\usepackage{hyperref}
\usepackage{bbold}
\usepackage{color}
\renewcommand{\d}[2]{\ensuremath{\frac{\text{d} #1}{\text{d} #2}}}

\newcommand{\pd}[2]{\ensuremath{\frac{\partial #1}{\partial #2}}}

\newcommand{\ket}[1]{\ensuremath{\left| #1 \right>}}

\renewcommand{\exp}[1]{\ensuremath{ \; \text{exp} \left( #1 \right) } }

\begin{document}

\preprint{APS/123-QED}

\title{Quantum dot  in interacting environments }

\author{Colin Rylands} 
\email{rylands@physics.rutgers.edu}
\author{Natan Andrei}
\email{natan@physics.rutgers.edu}
\affiliation{Department of Physics, Rutgers University
Piscataway, New Jersey 08854.
}

\date{\today}
              
 \begin{abstract}
 A quantum impurity  attached  to an interacting quantum wire   gives rise to an array of  of new phenomena.  Using Bethe Ansatz we solve exactly  models describing  two geometries  of 
 a quantum dot coupled to an interacting quantum wire:  a quantum dot that is (i) side-coupled and (ii)  embedded in a Luttinger liquid. We find the  eigenstates and determine the spectrum through the Bethe Ansatz equations. Using this we derive  exact expressions for the ground state dot occupation. The thermodynamics are then studied using the thermodynamics Bethe Ansatz equations. It is shown that at low energies the dot becomes fully hybridized and acts as a backscattering impurity or tunnel junction depending on the geometry and furthermore that the two geometries are related by changing the sign of the interactions. Although remaining strongly coupled for all values of the interaction in the wire, there exists competition between the tunneling and backscattering leading to a suppression or enhancement of the dot occupation depending on the sign of the bulk interactions.
  \end{abstract}

\maketitle
\section{Introduction}
\begin{figure}
\centering
\includegraphics[trim = 20mm 15mm 0mm 10mm, clip, width=0.5\textwidth]{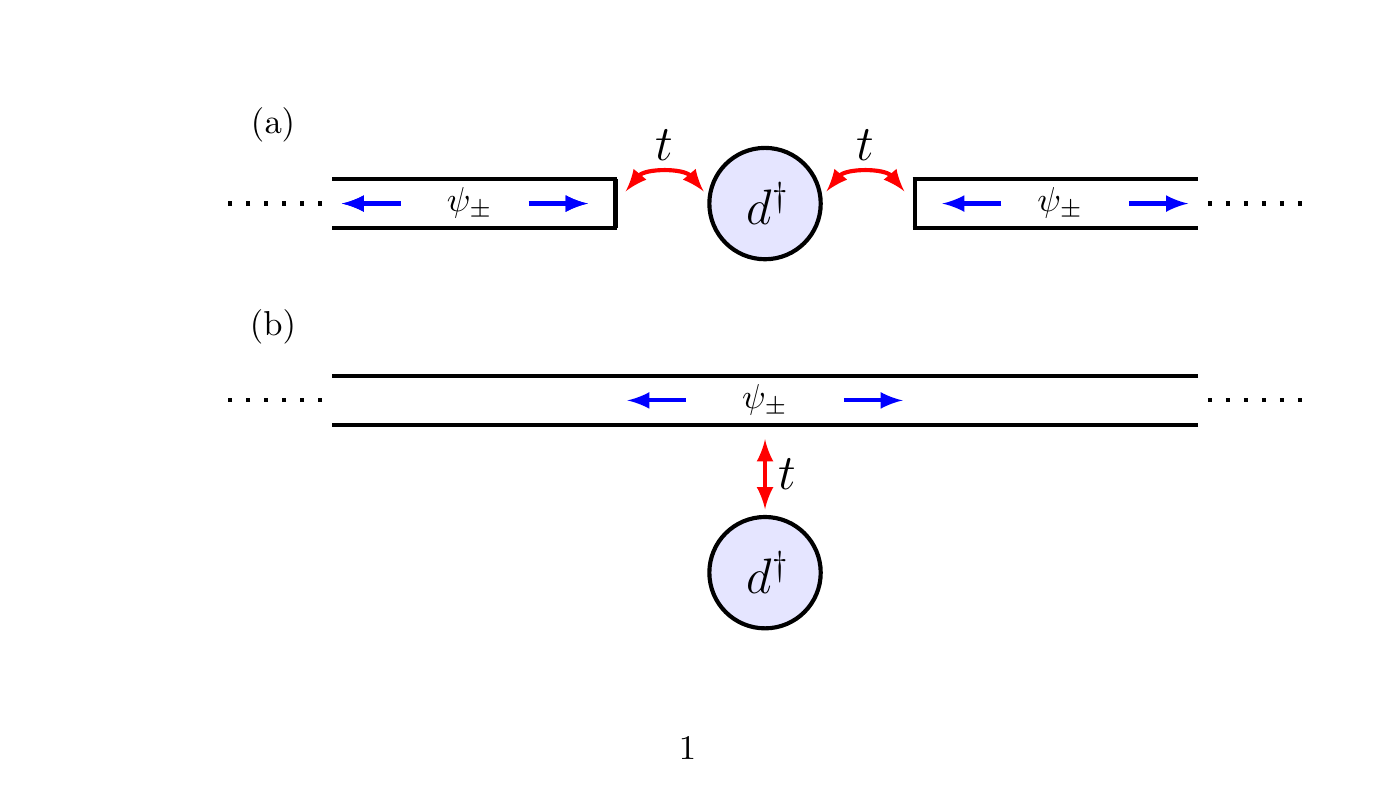}
\caption{We consider two geometries of Luttinger dot system; (a) embedded and (b) side-coupled. The embedded geometry also includes a Coulomb interaction between the dot and leads. Once unfolded the side-coupled and embedded geometries are the same but with the latter containing non local interactions \eqref{HLLem}}
\end{figure}
Coupling a quantum impurity to an interacting one dimensional lead produces some of the most striking phenomena of low dimensional physics. A simple backscattering  impurity is known to cause the wire to be split if the interactions are repulsive while a junction between two leads can lead to perfect conductance in the presence of attractive interactions \cite{KF}. More interesting still are scenarios in which the impurity has internal degrees of freedom. These allow for richer and more exotic phases to appear \cite{TG}. Among these, systems of quantum dots coupled to interacting leads have attracted much attention \cite{KF}\cite{FurNag}\cite{GogoKom}\cite{FurMat}\cite{Furusaki}\cite{GWB}\cite{Duality}\cite{Cap}\cite{Paata}\cite{Meden}\cite{Lerner}. The low energy description of the leads is typically given by Luttinger liquid theory which is the effective low energy description of a large number  of  interacting systems  \cite{haldane, TG}.  Here the  individual electrons are dissolved and the excitations are bosonic density modes. In contrast,  the relevant degrees of freedom on the dot are electronic. A competition ensues between the tunneling from lead to dot which is carried out by electrons and the energy cost of reconstituting an electron from the bosons in the lead. 

Such systems are readily achievable in many experimental settings allowing for confrontation of theory with experiment. Luttinger liquids provide the effective description of  carbon nano tubes \cite{LLCNT}\cite{JCNT}, fractional quantum Hall edges \cite{LLRMP}\cite{FQHLL}\cite{FQHLL2}, cold atomic gases  \cite{LLcold}\cite{ColdIRL}\cite{Jiang}\cite{CA} or $^4$He flowing through nano pores \cite{DelMae}\cite{Duc} to name but a few. Additionally they are known to describe tunneling processes in higher dimensional resistive leads \cite{Safi}\cite{Uni} and more generally are the archetype of a non-Fermi liquid. Luttinger liquid-quantum dot systems have successfully been realized in a number of experiments \cite{qpt}\cite{Meb}. These realize the embedded geometry, see Fig. 1(a) of a dot placed between two otherwise disconnected leads.  Measurement of the conductance has revealed interesting non-Fermi liquid scaling as well as Majorana physics.

Building on the work of \cite{ryl}\cite{CR2} we use Bethe Ansatz  to solve exactly Luttinger liquid-quantum dot systems in both  the embedded  (see Fig 1(a))  and  side coupled (see Fig 1(b)) geometries.  The exact solution shows that the spectra of the two geometries are related by changing the sign of the bulk interaction, a fact previously known through bosonization \cite{Duality}, and are described in terms of charge and chiral degrees of freedom.  At low energies we show that the dot becomes fully hybridized and acts as a backscattering impurity for the side-coupled model and as  a tunnel junction for the embedded system. This creates a competition between the charge and chiral degrees of freedom when the back scattering or tunnel junction is irrelevant,  leading to non Fermi liquid exponents in the ground state dot occupation. We then go on to study the finite temperature properties of the system deriving the Thermodynamic Bethe Ansatz equations and using this to obtain the finite temperature dot occupation.

This paper is organised as follows:  in section II we introduce the Hamiltonians and construct their exact eigenstates. We derive the exact spectrum of both systems through their Bethe Ansatz equations by means of the off diagonal Bethe Ansatz method (ODBA) \cite{ODBA}. In Section III we find the ground state of the system and derive the exact dot occupation. From this we extract the renormalization group picture of the system and find the leading relevant and irrelevant operators in section IV. The thermodynamics of the system are studied in section V where we find the free energy contribution from the dot and use it to obtain the dot occupation at finite temperature. In the final section we conclude.

\section{Models and eigenstates}

The systems we  consider consist of a  quantum dot attached  to an interacting lead,  a Luttinger liquid,  the attachment being either in   the embedded or the side-coupled geometry. The Hamiltonian of a Luttinger liquid is given by,
\begin{eqnarray}\nonumber
   H_{\text{LL}}&=&-i  \int  dx (\psi^\dagger_+ \partial_x 
\psi_+   -\psi^\dagger_- \partial_x \psi_-)
   \\\label{HLLsc}
   &&+ 4g \int   dx  \,\psi_+^\dag(x)\psi^\dag_-(x)\psi_-(x)\psi_+(x)
\end{eqnarray}
where $\psi^\dag_{\pm}$ are right and left moving fermions which interact with a point like interaction of strength $g$ \cite{TG}.  For the side-coupled geometry  we have $x\in [-L/2, L/2]$
while for the embedded geometry we take two Luttinger liquids restricted to  $x\in [-L/2, 0]$ and $x\in [0, L/2]$.
It is convenient to bring the two systems into similar form by unfolding the embedded geometry  in the standard way \cite{gogolin2004bosonization} to give,
\begin{eqnarray}\nonumber
   H^{\text{emb}}_{\text{LL}}&=&-i  \int  dx (\psi^\dagger_+ \partial_x 
\psi_+   -\psi^\dagger_- \partial_x \psi_-)
   \\\label{HLLem}
   &&+ 4g \sum_{\sigma=\pm}\int   dx  \,\psi_\sigma^\dag(x)\psi^\dag_\sigma(-x)\psi_\sigma(-x)\psi_\sigma(x)
\end{eqnarray}
The embedded system now consists of one branch of left-movers and one branch of right movers restricted to  $x\in [-L/2, L/2]$ but unlike the side-coupled system where the left and right fermions interact locally with each other, in the  embedded system after unfolding the interaction is  between particles of the same chirality and is non local.  Further, the spectrum being linear  a cut-off  needs to be imposed to render the energies finite.  We shall impose it on the particle momenta: $k \ge -\mathcal{D}$.
  All physical quantities are taken to be small compared with the cutoff and  at the end of the calculation we send $\mathcal{D}\to\infty$, to obtain universal results.

 The quantum dot is modelled by a resonant level with energy $\epsilon_0$ described by,
 \begin{eqnarray}\label{Hd}
  H_{\text{dot}}=\epsilon_0d^\dag d,
  \end{eqnarray}
 coupled to Luttinger liquid via a tunnelling term,
\begin{eqnarray}\label{Ht}
H_t&=& \frac{t}{2} (\psi^\dagger_+(0)+\psi^\dagger_-(0)) d +\text{h.c}
\end{eqnarray}
which mediates both forward and backscattering in the model, the latter changing left movers to right movers and vice versa. 
Furthermore in the embedded system we add a Coulomb interaction between the ends of the leads and the dot which is the same strength as the Luttinger interaction,
\begin{eqnarray}\label{Hc}
H_c=gd^\dag d\sum_{\sigma=\pm}\psi_{\sigma}^\dag(0)\psi_{\sigma}(0).
\end{eqnarray}
 Both energy scales in the dot Hamiltonian are small compared the the cut-off,
   $\epsilon_0,\Gamma\ll \mathcal{D}$, where $\Gamma= t^2$ is the level width.

 We shall determine the spectrum and the full set of  exact eigenstates  of both Hamiltonians,  $H^\text{sc}=H_{LL}+H_t+H_{\text{dot}}$ and $H^\text{emb}=H^\text{emb}_{LL}+H_t+H_{\text{dot}}+H_c$, using the Bethe Ansatz approach, and then  proceed to the ground state (T=0)  and thermodynamic  ($T>0$)  properties.  The Bethe Ansatz method we employ here is distinct from that which has been typically used for quantum impurity models \cite{4lectures}\cite{TWAKM}. As the problem contains both forward and back scattering we must formulate it in an in-out basis  with the configuration space being partitioned in regions labelled by both the order of the particles and by their  closeness to the origin. The large degeneracy present in the bulk system due to the linear derivative is then used to find  a consistent set of wave functions \cite{ryl}. We illustrate this by explicitly constructing the one and two particle eigenstates from which we can determine the $N$-particle states.
 \begin{figure}
\centering
\includegraphics[trim = 10mm 18mm 0mm 0mm, clip, width=0.22\textwidth]{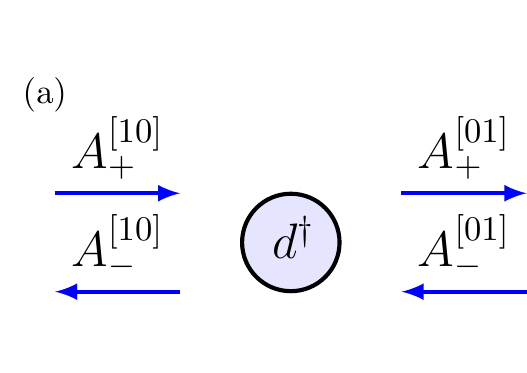}
\includegraphics[trim = 0mm 18mm 0mm 5mm, clip, width=0.22\textwidth]{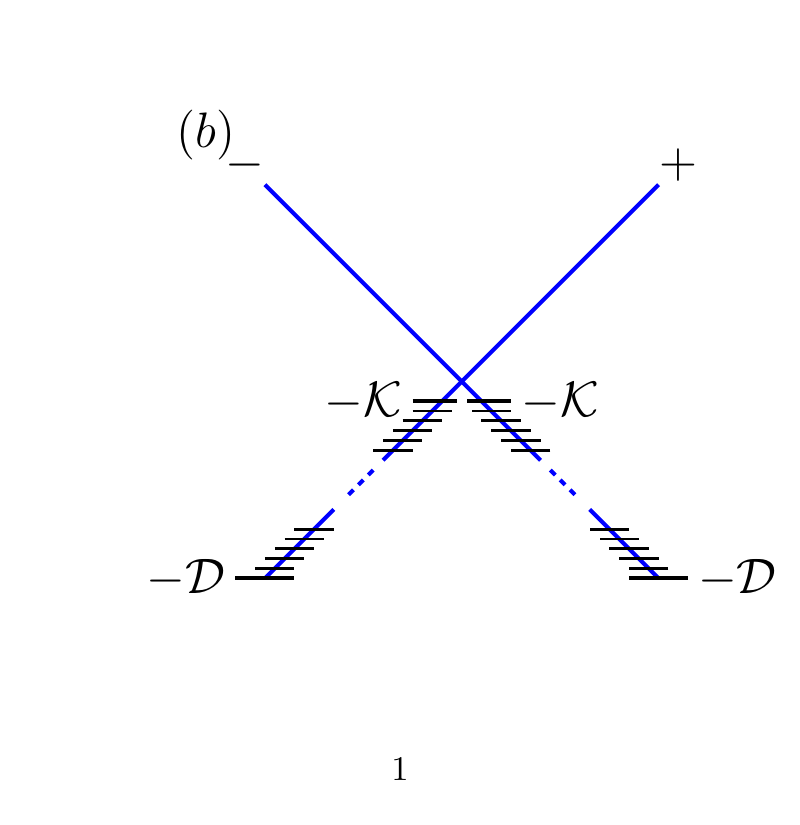}
\caption{(a) The single particle wavefunction  given by \eqref{Sin} is depicted. Particles are either incoming on the left or right with amplitudes  $A^{[10]}_+,A^{[01]}_-$ or outgoing on the left or right with amplitudes  $A^{[10]}_-,A^{[01]}_+$. (b) The linear derivative requires that we cutoff the bottom of the Dirac sea so that $k>-\mathcal{D}$ which we will take to infinity in the end. When the rapidity notation is used the dot energy acts as a local chemical potential and in the ground state levels are filled up to $-\mathcal{K}=-\mathcal{D}e^{-B/2}$, with $B=B(\epsilon_0)$.}
\end{figure}
 
 After the unfolding procedure the two systems differ only in the two particle interaction meaning the single particle eigenstates are the same in both models. The tunnelling to and from the dot takes place at the origin hence we may expand the wavefunction in plane waves on either side of it,  the most general form for the single particle state of energy $E=k$  being,
 \begin{eqnarray}\nonumber
\sum_{\sigma=\pm} \int e^{\sigma ikx}\left[\theta(-x)A^{[10]}_\sigma+\theta(-x)A_\sigma^{[01]}\right]\psi^\dag_\sigma(x)\ket{0}\\\label{Sin}
+Bd^\dag\ket{0},
 \end{eqnarray}
where $\theta(\pm x)$ are Heaviside functions.
The amplitudes $A^{[10]}_+$ and $A^{[01]}_-$ are those of an incoming particle and are related to the outgoing amplitudes $A^{[10]}_-$ and $A^{[01]}_+$ (see Fig. 2(a)) by the bare single particle S-matrix - $S$, which takes an incoming particle to an outgoing one.  Trading in the particle momentum $k$ for the rapidity variable  $z$, defined as
$
 k-\epsilon_0=\mathcal{D}e^{z/2},
 $ 
we have,   
\begin{eqnarray}
\begin{pmatrix}
A^{[01]}_+\\
A^{[10]}_-
\end{pmatrix}=S(z)
\begin{pmatrix}
A^{[10]}_+\\
A^{[01]}_-
\end{pmatrix}\\\label{s}
S(z)=
\begin{pmatrix}
\frac{e^{z/2}}{e^{z/2}+ie^c} && \frac{-ie^c}{e^{z/2}+ie^c}\\
\frac{-ie^{c}}{e^{z/2}+ie^c} &&\frac{e^{z/2}}{e^{z/2}+ie^c}
\end{pmatrix}
\end{eqnarray}
with $e^c=\Gamma/\mathcal{D}$. In addition the dot amplitude $B$ is 
\begin{eqnarray}
B=\sum_{\sigma=\pm}\frac{1 }{2}e^{(c-z)/2}\left(A_{\sigma}^{[10]}+A^{[01]}_\sigma\right).
\end{eqnarray}
From here periodic boundary conditions can be imposed $\psi_\pm^\dag(-L/2)=\psi_\pm^\dag(L/2)$ resulting in 
\begin{eqnarray}
e^{-i\mathcal{D}e^{z/2}L-i\epsilon_0L}\begin{pmatrix}
A^{[10]}_+\\
A^{[01]}_-
\end{pmatrix}=S(z)
\begin{pmatrix}
A^{[10]}_+\\
A^{[01]}_- 
\end{pmatrix}
\end{eqnarray}
which can then be solved for the allowed values of the rapidity $z$.

We now proceed to the two particle case wherein the bulk interaction $g$ enters differently in both models.  We shall first  consider the side-coupled model and discuss the embedded model subsequently. Since the two particle interaction is point-like as is the tunnelling to the dot we may divide configuration space into regions such that the interactions only occur at the boundary between two regions. Therefore away from these boundaries we write the wavefunction as a sum over plane waves. For two particles we require 8 regions which are specified not only by the ordering of the particle positions $x_1$, $x_2$ and  the impurity but also according to which position is closer to the origin. For example if $x_1$ is to the left of the impurity, $x_2$ to its right  with $x_2$ closer  to the impurity then the amplitude in this region is denoted $A^{[102B]}_{\sigma_1\sigma_2}$, $\sigma_j=\pm$ being the chirality of the particle at $x_j$. The region in which $x_1$ is closer is denoted  $A^{[102A]}_{\sigma_1\sigma_2}$.  The consequence for the wavefunction is that we include Heaviside functions $\theta (x_Q)$ which have support only in the region $Q$, e.g $\theta(x_{[102B]})=\theta(x_2)\theta(-x_1)\theta (-x_1-x_2)$ and $ \theta(x_{[102A]})=\theta(x_2)\theta(-x_1)\theta (x_1+x_2)$.  The most general two particle state with energy $E=k_1+k_2=\sum_{j=1}^2\mathcal{D}e^{z_j/2}+2\epsilon_0$ is  therefore 
\begin{eqnarray}\nonumber
\ket{E}&=&\sum_Q\sum_{\sigma_1,\sigma_2=\pm}\int\theta (x_Q)A_{\sigma_1\sigma_2}^{Q}\prod_j^2e^{i\sigma_jk_j x_j} \psi^\dag_{\sigma_j}(x_j)\ket{0}\\\label{2part}
&+&\sum_{\sigma=\pm}\int\left[\theta (-x)B_{\sigma}^{[10]}+\theta (x)B_{\sigma}^{[01]}\right] \psi^\dag_{\sigma}(x)d^\dag\ket{0}.
\end{eqnarray}
The amplitudes $A_{\sigma_1\sigma_2}^{Q}$ are related to each other by S-matrices which are fixed by the Hamiltonian and in turn fix $B^{[10]}_\pm$ and $B^{[01]}_\pm$. To define these S-matrices we form column vectors of the amplitudes,
\begin{eqnarray}
\centering\nonumber
&&\vec{A}_1=\begin{pmatrix}
A_{++}^{[120B]}\\
A_{+-}^{[102B]}\\
A_{-+}^{[201B]}\\
A_{--}^{[021B]}
\end{pmatrix}~\vec{A}_2=\begin{pmatrix}
A_{++}^{[210A]}\\
A_{+-}^{[102A]}\\
A_{-+}^{[201A]}\\
A_{--}^{[012A]}
\end{pmatrix}~\vec{A}_3=\begin{pmatrix}
A_{++}^{[201A]}\\
A_{+-}^{[012A]}\\
A_{-+}^{[210A]}\\
A_{--}^{[102A]}
\end{pmatrix}\\\nonumber
&&\vec{A}_4=\begin{pmatrix}
A_{++}^{[201B]}\\
A_{+-}^{[021B]}\\
A_{-+}^{[120B]}\\
A_{--}^{[102B]}
\end{pmatrix}
~\vec{A}_5=\begin{pmatrix}
A_{++}^{[021B]}\\
A_{+-}^{[201B]}\\
A_{-+}^{[102B]}\\
A_{--}^{[120B]}
\end{pmatrix}
~\vec{A}_6=\begin{pmatrix}
A_{++}^{[012A]}\\
A_{+-}^{[201A]}\\
A_{-+}^{[102A]}\\
A_{--}^{[210A]}
\end{pmatrix}
\\\label{amps}&&~~~~~~\vec{A}_7=\begin{pmatrix}
A_{++}^{[102A]}\\
A_{+-}^{[210A]}\\
A_{-+}^{[012A]}\\
A_{--}^{[201A]}
\end{pmatrix}
~~~~~~~\vec{A}_8=\begin{pmatrix}
A_{++}^{[102B]}\\
A_{+-}^{[120B]}\\
A_{-+}^{[021B]}\\
A_{--}^{[201B]}
\end{pmatrix}
\end{eqnarray} 
which have the following interpretation:  $\vec{A}_1$ ($\vec{A}_2$) are the amplitudes where both particles are incident on the impurity but particle 2 (1) is closer,  $\vec{A}_5$ ($\vec{A}_6$) are the amplitudes in which both particles are outgoing with particle 2 (1) closer to the impurity, $\vec{A}_8$ ($\vec{A}_3$) describes  particle 2 (1)  having scattered off the impurity and is still closer to the impurity than 1 (2) while $\vec{A}_7$  ($\vec{A}_4$ )  also describes particle 2 (1)  having scattered but with 1 (2) is closer. $\vec{A}_1$ and $\vec{A}_8$ are explicitly depicted in Fig. 3. 
  \begin{figure}
\centering
\includegraphics[trim = 10mm 10mm 0mm 0mm, clip, width=0.5\textwidth]{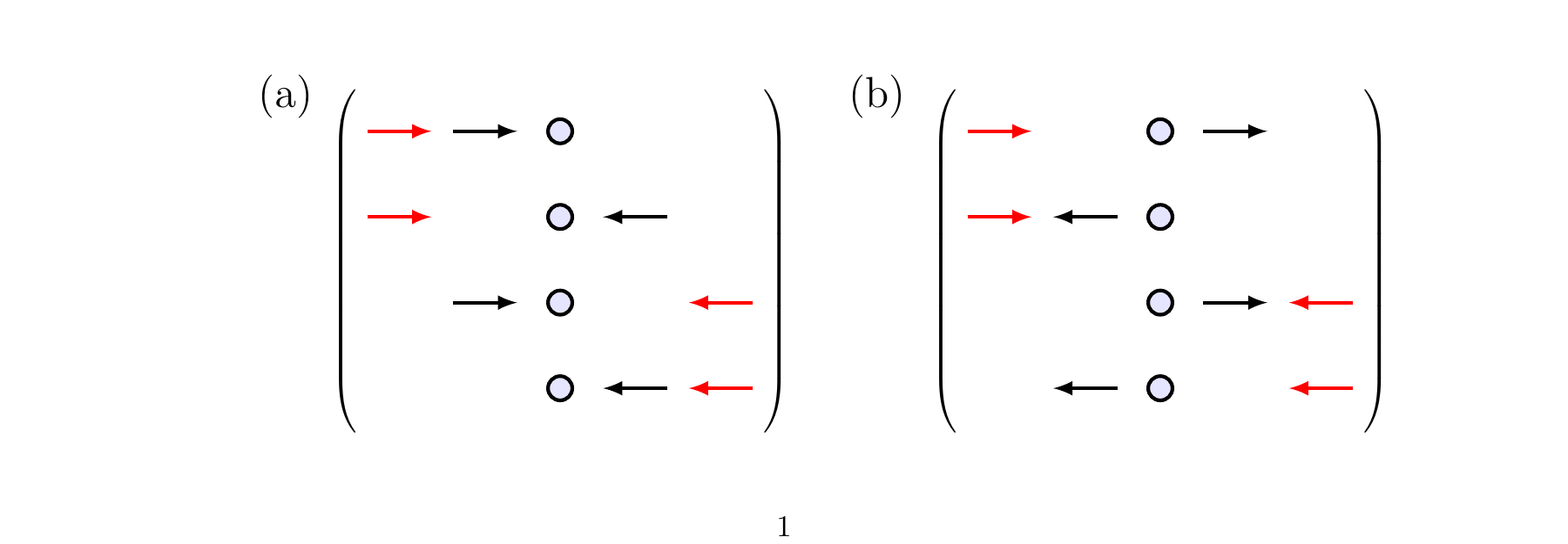}
\caption{(Color online) The amplitudes in the two particle wavefunction are arranged into $8$ vectors given by \eqref{amps} and according to whether the particles are incoming or outgoing as well as their ordering with respect to the impurity. (a) The amplitudes in $\vec{A}_1$ consist of both particles incoming but particle 2 (black) closer to the impurity than particle 1 (red). (b) The amplitudes in $\vec{A}_8$ consist of particle two outgoing. These vectors are related by $S^{20}(z_2)$.}
\end{figure}
 After applying the Hamiltonian to \eqref{2part} we find that it is an eigenstate provided,
 \begin{eqnarray}\label{S1}
\vec{A}_8=S^{20}(z_2)\vec{A}_1,~&~&~\vec{A}_3=S^{10}(z_1)\vec{A}_2,\\
\vec{A}_5=S^{20}(z_2)\vec{A}_4,~&~&~\vec{A}_6=S^{10}(z_1)\vec{A}_7,\\
\vec{A}_7=S^{12}\vec{A}_8,~&~ &~\vec{A}_4=S^{12}\vec{A}_3,\\
\vec{A}_2=W^{12}(z_2-z_1)\vec{A}_1,&~&\vec{A}_6=W^{12}(z_2-z_1)\vec{A}_5.
\end{eqnarray}
The S-matrices $S^{20}$ and $S^{10}$ which take a particle past the impurity, i.e. from incoming to outgoing are 
\begin{eqnarray}
S^{20}(z_2)=S(z_2)\otimes \mathbb{1},~~~S^{10}(z_1)=\mathbb{1}\otimes S(z_1),\label{S}
\end{eqnarray}
with $S(z)$ the same as in the single particle state \eqref{s}, the  S-matrix $S^{12}$ scatters an incoming particle past an outgoing particle and is 
\begin{eqnarray}
S^{12}=\begin{pmatrix}
1&&0&&0&&0\\
0&&e^{i\phi}&&0&&0\\
0&&0&&e^{i\phi}&&0\\
0&&0&&0&&1
\end{pmatrix}.
\end{eqnarray}
where $\phi=-2\arctan{(g)}$ encodes the bulk interaction and $W^{12}(z_2-z_1)$ which scatters an incoming (outgoing) particle past another incoming (outgoing) particle is given by
\begin{eqnarray}
W^{12}(z)=
\begin{pmatrix}
1&&0&&0&&0\\
0&&\frac{\sinh{\frac{1}{2}(z)}}{\sinh{\frac{1}{2}(z-2i\phi)}}&&\frac{-\sinh{i\phi}}{\sinh{\frac{1}{2}(z-2i\phi)}}&&0\\
0&&\frac{-\sinh{i\phi}}{\sinh{\frac{1}{2}(z-2i\phi)}}&&\frac{\sinh{\frac{1}{2}(z)}}{\sinh{\frac{1}{2}(z-2i\phi)}}&&0\\
0&&0&&0&&1
\end{pmatrix}.
\end{eqnarray}
In addition the dot amplitudes are given by
\begin{eqnarray}\nonumber
B^{[10]}_\pm=\frac{1}{2}e^{(c-z_2)/2}\sum_{\sigma=\pm}\left[A^{[210A]}_{\sigma\pm}+A^{[201A]}_{\sigma\pm}\right]\\\label{B1}
-\frac{1}{2}e^{(c-z_1)/2}\sum_{\sigma=\pm}\left[A^{[120B]}_{\pm\sigma}+A^{[102B]}_{\pm\sigma}\right],\\\nonumber
B^{[01]}_\pm=\frac{1}{2}e^{(c-z_2)/2}\sum_{\sigma=\pm}\left[A^{[102A]}_{\sigma\pm}+A^{[012A]}_{\sigma\pm}\right]\\\label{B2}
-\frac{1}{2}e^{(c-z_1)/2}\sum_{\sigma=\pm}\left[A^{[201B]}_{\pm\sigma}+A^{[021B]}_{\pm\sigma}\right].
\end{eqnarray}
Inserting these expressions for the amplitudes into \eqref{2part} we get the two particle eigenstate of the side-coupled model. Since all amplitudes are generated from $\vec{A}_1$ by successive application of the various S-matrices, as depicted in Fig. 4,  there are  two ways to obtain each $\vec{A}_j$ both of which must be equivalent for the construction to be consistent.  This consistency imposes that the S-matrices satisfy a generalised Yang Baxter equation which takes the form of the reflection equation 
\begin{equation}\label{RE}
S^{20}S^{12}S^{10}W^{12}=W^{12}S^{10}S^{12}S^{20}
\end{equation}
\begin{figure}
\includegraphics[trim=30 30 30 30,width=.5\textwidth]{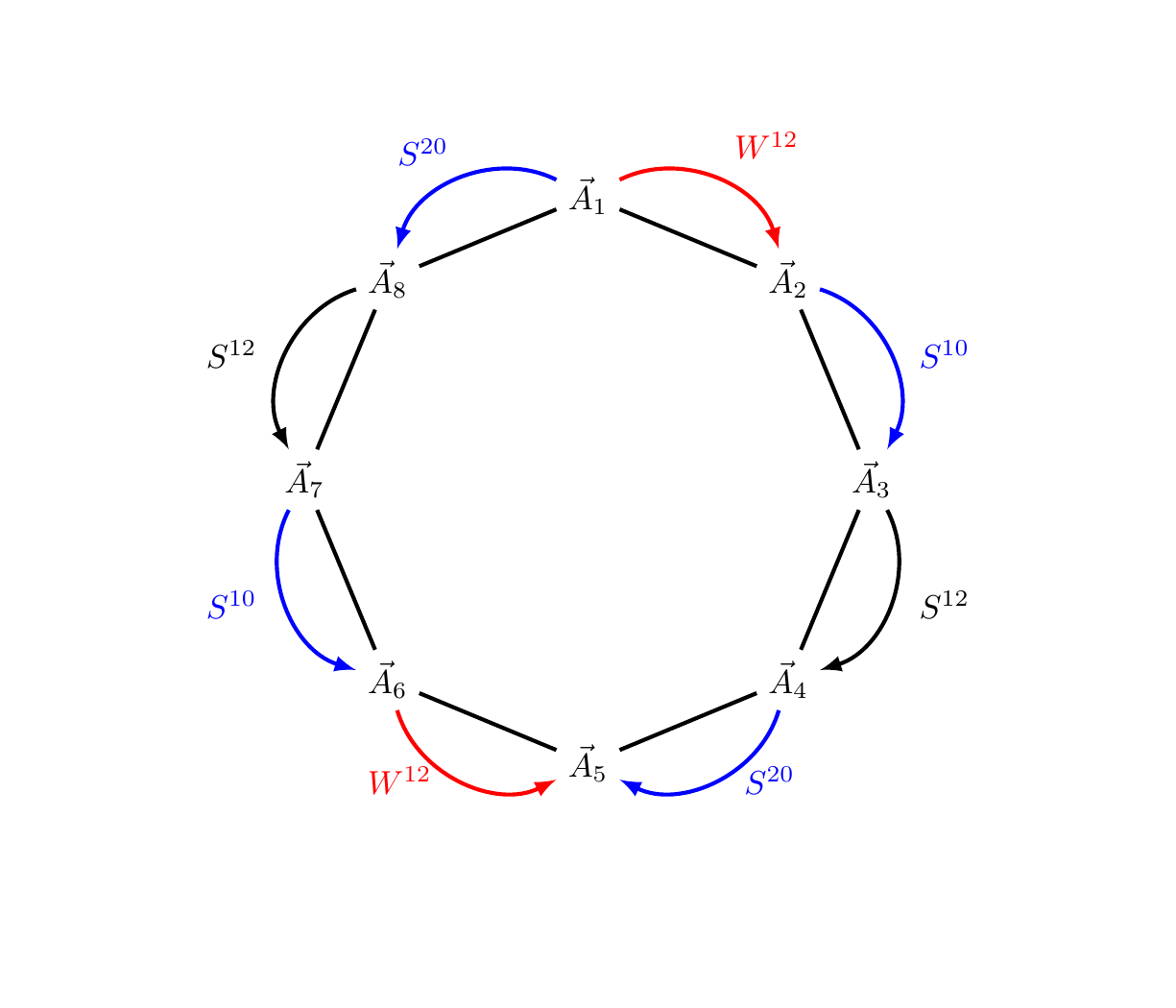}\label{fRE}
\caption{(Color Online)The amplitudes are related by applying the operators as in \eqref{S1} and depicted here. For consistency we require the amplitudes obtained by proceeding clockwise or counter-clockwise are the same resulting in \eqref{RE}.}
\end{figure} 
which can be checked to hold by substitution. 

It is important to note that while no interaction between two incoming (outgoing) particles is present in the Hamiltonian, $W^{12}$ is introduced in order to obtain the correct eigenstates and satisfy the generalised Yang-Baxter consistency conditions.  To do so we exploit the freedom to introduce discontinuities of the form $\theta(\pm(x_1-x_2))$   into the the part of the wave function that describes two right movers or two left movers  (or $\theta(\pm(x_1+ x_2))$   into the the part of the wave function that describes one left mover and one right mover). The kinetic term in the Hamiltonian referring to these particles is of the form $\pm i(\partial_{x_{1}} +i\partial_{x_{2}})$ (or $\pm i(\partial_{x_{1}} -i\partial_{x_{2}})$)  and  vanishes when acting on these discontinuities. This freedom arises from the linear spectrum that brings about a infinite degeneracy of the energy levels, the level $k_1+k_2$ being degenerate with $(k_1+q) + (k_2-q)$ for any $q$. The introduction of the discontinuities corresponds then to the correct choice of basis states in this degenerate subspace from which the perturbation can be turned on, as we are instructed to do carrying out perturbation theory from a degenerate level. For more detail see \cite{4lectures}.

We can then go on to impose periodic boundary conditions giving
\begin{eqnarray}
e^{-ik_1L}\vec{A}_1&=&S^{12}S^{10}W^{12}\vec{A}_1\\
e^{-ik_2L}W^{12}\vec{A}_1&=&S^{12}S^{20}\vec{A}_1
\end{eqnarray}
which can be solved to determine $z_{1,2}$.

The eigenstates for higher particle number are constructed similarly and the $N$ particle state with energy $E=\sum_{j=1}^N k_j=\sum_{j=1}^N\mathcal{D}e^{z_j/2}+N\epsilon_0$ is,

\begin{eqnarray}\nonumber
\ket{E}&=&\sum_Q\sum_{\vec{\sigma}}\int\theta (x_Q)A_{\vec{\sigma}}^{Q}\prod_j^Ne^{i\sigma_jk_j x_j} \psi^\dag_{\sigma_j}(x_j)\ket{0}\\\label{Npart}
&+&\sum'_P\sum'_{\vec{\sigma}}\int\theta (x_P)B_{\vec{\sigma}}^{P}\prod'_je^{i\sigma_jk_j x_j} \psi^\dag_{\sigma_j}(x_j)d^\dag\ket{0}
\end{eqnarray}
Here $\theta(x_Q)$ are Heaviside  functions which partition configuration space into $2^NN!$ regions. As before $Q$ are labelled by the ordering of the N particles as well as according to which particle is closest to the origin while  $\vec{\sigma}=(\sigma_1,\dots,\sigma_N)$ with $\sigma_j=\pm$. In the second line the primed sums indicate that one particle is removed - being on the dot - and the sums are over the remaining $N-1$ particle system. Just as in the two particle case the amplitudes are related to each other via S-matrices which act on the $2^N$ dimensional space 
\begin{eqnarray}\label{Sj}
S^{j0}&=&S_j(z_j)\otimes_{k\neq j}\mathbb{1},\\\label{Sij}
S^{ij}&=&\begin{pmatrix}
1&&0&&0&&0\\
0&&e^{i\phi}&&0&&0\\
0&&0&&e^{i\phi}&&0\\
0&&0&&0&&1
\end{pmatrix}_{ij}\otimes_{k\neq i,j}\mathbb{1},~~\\\nonumber
~~W^{ij}\!&=&\!\begin{pmatrix}
1&&0&&0&&0\\
0&&\frac{\sinh{\frac{1}{2}(z_j-z_i)}}{\sinh{\frac{1}{2}(z_j-z_i-2i\phi)}}&&\frac{-\sinh{i\phi}}{\sinh{\frac{1}{2}(z_j-z_i-2i\phi)}}&&0\\
0&&\frac{-\sinh{i\phi}}{\sinh{\frac{1}{2}(z_j-z_i-2i\phi)}}&&\frac{\sinh{\frac{1}{2}(z_j-z_i)}}{\sinh{\frac{1}{2}(z_j-z_i-2i\phi)}}&&0\\
0&&0&&0&&1
\end{pmatrix}_{ij}\\\label{Wij}
&&\otimes_{k\neq i,j}\mathbb{1}.
\end{eqnarray}
where the subscripts denote which particle spaces the operators act upon. In order for this wavefunction to be consistent it must satisfy the following Yang-Baxter and reflection equations,
\begin{eqnarray}\label{REN}
S^{k0}S^{jk}S^{j0}W^{jk}&=&W^{jk}S^{j0}S^{jk}S^{k0}\\\label{YB1}
W^{jk}W^{jl}W^{kl}&=&W^{kl}W^{jl}W^{jk}\\\label{YB2}
W^{jk}S^{jl}S^{kl}&=&S^{kl}S^{jl}W^{jk}.
\end{eqnarray} 
The first of these being the generalisation to $N$ particles of \eqref{RE} while the remaining two come from the consistency of the wavefunction away from the dot. These  are indeed satisfied by \eqref{Sj},\eqref{Sij} and \eqref{Wij} which is a sufficient condition for the consistency of the wave function \citep{ZinnJustin}. The expressions for $B_{\vec{\sigma}}^{P}$ in terms $A_{\vec{\sigma}}^{Q}$ can also be found and are straightforward generalisations of  \eqref{B1} and \eqref{B2}. Therefore we have successfully constructed the $N$ particle eigenstates of the side-coupled model. 

The spectrum  can then be determined by imposing periodic boundary conditions $\psi^\dag_\pm(-L/2)=\psi^\dag_\pm(L/2)$. As we are interested in studying properties of the dot in the thermodynamic limit the type of boundary condition imposed at $x=\pm L/2$ will not affect the result. This results in an eigenvalue problem which determines the $k_j$ through
\begin{eqnarray}\label{PBC}
&& e^{-ik_jL}A_{\sigma_1\dots\sigma_N}=\left(Z_j\right)^{\sigma'_1\dots\sigma'_N}_{\sigma_1\dots\sigma_N}A_{\sigma'_1\dots\sigma'_N}\\
 &&Z_j=W^{j-1j}.. W^{1j}S^{1j}.. S^{jN}S^{j0}W^{jN}.. W^{jj+1}
\end{eqnarray} 
where the matrix $Z_j$ takes the $j$th particle past all others and past the impurity. By using \eqref{RE}, \eqref{YB1} and \eqref{YB2} one can show that the $Z_j$ commute with each other $[Z_j,Z_k]=0$ $\forall j,k$. They are therefore simultaneously diagonalisable and the spectrum of the side-coupled model is determined by the eigenvalues of the $Z_j$ operators. Before obtaining these we return to constructing the eigenstates of the embedded model.

For the embedded impurity model we note that the unfolding procedure carried out previously allows us to  construct its eigenstates in the same manner as we did for  the side-coupled model. The $N$ particle eigenstate is of the same form as \eqref{Npart} but owing to the different bulk interaction in \eqref{HLLem} the two particle S-matrices are
\begin{eqnarray}
S^{ij}_\text{emb}&=&\begin{pmatrix}
e^{i\phi}&&0&&0&&0\\
0&&1&&0&&0\\
0&&0&&1&&0\\
0&&0&&0&&e^{i\phi}
\end{pmatrix}_{ij}\otimes_{k\neq i,j}\mathbb{1},~~\\\nonumber
~~W^{ij}_\text{emb}\!&=&\!\begin{pmatrix}
1&&0&&0&&0\\
0&&\frac{\sinh{\frac{1}{2}(z_j-z_i)}}{\sinh{\frac{1}{2}(z_j-z_i+2i\phi)}}&&\frac{\sinh{i\phi}}{\sinh{\frac{1}{2}(z_j-z_i+2i\phi)}}&&0\\
0&&\frac{\sinh{i\phi}}{\sinh{\frac{1}{2}(z_j-z_i+2i\phi)}}&&\frac{\sinh{\frac{1}{2}(z_j-z_i)}}{\sinh{\frac{1}{2}(z_j-z_i+2i\phi)}}&&0\\
0&&0&&0&&1
\end{pmatrix}_{ij}\\\label{Wij}
&&\otimes_{k\neq i,j}\mathbb{1}.
\end{eqnarray}
and the single particle S-matrices $S^{j0}$ the same as \eqref{Sj}. The inclusion of the Coulomb term \eqref{Hc} is essential for this and in its absence the model is not integrable. 

 Imposing  the boundary condition $\psi_\pm(-L/2)=e^{i\phi}\psi^\dag_\pm(L/2)$ we have another eigenvalue problem,
 \begin{eqnarray}
&& e^{-ik_jL}A_{\sigma_1\dots\sigma_N}=\left(Z^{\text{emb}}_j\right)^{\sigma'_1\dots\sigma'_N}_{\sigma_1\dots\sigma_N}A_{\sigma'_1\dots\sigma'_N}
\end{eqnarray}
where $Z_j^\text{emb}$ is defined similarly to $Z_j$ in \eqref{PBC} but using $W^{ij}_\text{emb}$ and $S^{ij}_\text{emb}$ and is related to $Z_j$ by
\begin{eqnarray}
Z_j^\text{emb}= Z_j|_{\phi\to-\phi}.
\end{eqnarray}
Therefore, the spectrum of the embedded model is obtained from the side-coupled model by changing the sign of the interaction, $\phi\to-\phi$. 

We can replace the bare phase shift $\phi$ by the universal Luttinger liquid parameter $K$ using 
\cite{ryl} \cite{CR2}
\begin{eqnarray}\label{K}
K=\begin{cases}
1+\frac{\phi}{\pi} &\text{side-coupled}\\
\frac{1}{1-\frac{\phi}{\pi}}& \text{embedded}
\end{cases}
\end{eqnarray}
meaning that in the thermodynamic limit the two models are related by taking $K\to 1/K$ which recovers the duality shown by bosonization \cite{Duality}. In the subsequent sections all calculations will be done for the side-coupled model the results of which can then be translated to the embedded model by taking $K\to 1/K$. Note that as $\phi$ is a phase shift and restricted to $[-\pi,\pi]$ we see that the side-coupled system may realize values of $K\in [0,2]$ whereas the embedded system has $K\in[1/2,\infty]$. 

\section{ Derivation of the Bethe Ansatz equations}

Our task now is to determine the eigenvalues of $Z_j$. To this end  we note that $W^{ij}$ is actually the R-matrix of the XXZ model, and further that $Z_j$ takes the form of the transfer matrix of an inhomogeneous open XXZ model \cite{Sklyannin}. The problem of diagonalising this operator has recently been achieved by means of the "Off Diagonal Bethe Ansatz" \cite{ODBA}. Inserting these results into \eqref{PBC} and simplifying the resulting equations using $e^c\ll 1$ in the same manner as in \cite{ryl} we obtain the Bethe equations for the side-coupled model 
\begin{eqnarray}\nonumber
e^{-i\mathcal{D}e^{z_\alpha/2}L}=e^{iN\phi/2+i\epsilon_0L}\left[\frac{e^{z_\alpha/2}-ie^c}{e^{z_\alpha/2}+ie^c}\right]^\frac{1}{2}\quad\quad\quad\quad\\\label{BAE1}
\times\prod^{N/2}_k\frac{\sinh{(\frac{1}{2}(z_\alpha-\lambda_k-i\phi))}}{\sinh{(\frac{1}{2}(z_\alpha-\lambda_k+i\phi))}}\quad\\\nonumber
\prod^{N}_\alpha\frac{\sinh{(\frac{1}{2}(\lambda_j-z_\alpha+i\phi))}}{\sinh{(\frac{1}{2}(\lambda_j-z_\alpha-i\phi))}}=-\left[\frac{\cosh{(\frac{1}{2}(\lambda_j-2c+i\phi))}}{\cosh{(\frac{1}{2}(\lambda_j-2c-i\phi))}}\right]^\frac{1}{2}\\\label{BAE2}\times\prod^{N/2}_k\frac{\sinh{(\frac{1}{2}(\lambda_j-\lambda_k+2i\phi))}}{\sinh{(\frac{1}{2}(\lambda_j-\lambda_k-2i\phi))}}\quad.
\end{eqnarray} 
where the parameters $\lambda_j$ describe the chiral degrees of freedom, $z_\alpha$ describe the charge degrees of freedom and the energy of the system is 
\begin{eqnarray}\label{E}
E=\sum_\alpha\mathcal{D}e^{z_\alpha/2}+N\epsilon_0.
\end{eqnarray}   
The solution of \eqref{BAE1}\eqref{BAE2} along with \eqref{E} give the exact energies of the system. 

\section{Ground state dot occupation}
Having obtained the Bethe equations governing the system we can now construct the ground state. To do this we first  must fill the empty Dirac sea with negative energy particles from the cutoff, $-\mathcal{D}$ up to some level determined by minimisation of the energy  (and  depending on  $\epsilon_0$, see Fig. 2(b)). After this the thermodynamic limit $N,L\to \infty$ is taken holding the density $D=N/L$ fixed and finally we take the universal limit by removing the cutoff $\mathcal{D}\to\infty$ while holding some other scale, which has been generated by the model, fixed. We will see below that this scale is the level width $\Gamma$. Once the ground state has been found we will use it to derive exact expressions for the occupation of the dot, $n_d=\left<d^\dag d\right>$ as a function of $\epsilon_0$.

The form of the possible negative energy states entering the ground state depends upon the value of $K$, whether it is greater or less than $1$ and so the ground state must be constructed separately in each case. Nevertheless we will find a single expression for the dot occupation valid in both regimes.
\subsection{$K>1$}
 We begin with  $\phi\in[0,\pi]$  which corresponds to $K\in[1,2]$. Here the ground state consists of so-called 2-strings \cite{Takahashi} wherein the rapidities form complex conjugate pairs with their real part coinciding with one of the chiral variables,
 \begin{eqnarray}
 z_j=z^*_{N+1-j}=\lambda_j+2\pi i+i\phi.
 \end{eqnarray}
with each pair having bare energy $ -2\cos{(\phi/2)}\mathcal{D}e^{\lambda_j}$, see Fig. 7.

Inserting these expressions  into \eqref{BAE1} and \eqref{BAE2} we obtain equations for  the real parts of the pairs, $\lambda_j$. In the thermodynamic limit we are not interested in the solutions per se, but in their  distribution, $$\rho_1(\lambda_j)=\frac{1}{L(\lambda_{j}-\lambda_{j-1})}$$ on the real line. The distribution has contributions from the bulk as well as from an $\mathcal{O}(1/L)$ term from the dot,  allowing  us to write it as $\rho_1(\lambda)=\rho^{\text{b}}_1(\lambda)+\frac{1}{L}\rho^\text{d}_1(\lambda)$.  The dot occupation is then given as,
 \begin{eqnarray}\label{ni}
n_d=2\int\rho^\text{d}_1(\lambda).
\end{eqnarray} 
The factor of $2$ appears here as each $\lambda$ corresponds to a pair of rapidities. These distributions,  $\rho^{\text{b}}_1(\lambda), \frac{1}{L}\rho^\text{d}_1(\lambda)$ are determined by the Bethe equations in their continuous form which for the bulk part is,
\begin{eqnarray}
\frac{\cos{\phi/2}}{2\pi}\mathcal{D}e^{\lambda/2}&=&\rho^\text{b}_1(\lambda)+\int^\infty_{-B}a_2(\lambda-\mu)\rho^\text{b}_1(\mu)\\
a_2(x)&=&\frac{i}{2\pi}\d{}{x}\log{\frac{\sinh{(\frac{1}{2}(x-ni\phi))}}{\sinh{(\frac{1}{2}(x+ni\phi))}}}
\end{eqnarray}
where $B=B(\epsilon_0)$ is the $\lambda$ value of the highest filled level. When the dot energy vanishes we have that $B(0)=\infty$ and bulk distribution is found to be 
\begin{eqnarray}
\rho_1^\text{b}(\lambda)=\frac{\mathcal{D}e^{\lambda/2}}{4\pi\cos{(\phi/2)}}
\end{eqnarray}
with the bulk part of the ground state energy being
\begin{eqnarray}
E_0=-\int_{-\infty}^\infty 2\cos{(\phi/2)}\mathcal{D}e^{\lambda/2}\rho_1^\text{b}(\lambda).
\end{eqnarray} 
To confirm this is indeed the ground state one can introduce excitations and check the energy is increased, the simplest type of which consists of adding holes to the distribution. As is typical for Bethe ansatz models, the energy of a hole turns out to be proportional to the ground state distribution i.e. a hole at $\lambda=\lambda^h$  has energy  $\varepsilon^h(\lambda^h)=4\pi\rho_1^\text{b}(\lambda^h)$, increasing the energy. The other excitations consist of breaking up a pair and placing them above the Fermi surface such they have real rapidity. Each particle then has energy $\varepsilon^p(z)=2\mathcal{D}e^{z/2}$ in addition to the hole introduced in the $\rho_1(\lambda)$ distribution.

When $\epsilon_0\neq 0$ the additional term  in the energy (see \eqref{E}) needs to be balanced by the addition of holes to the ground state with rapidities starting at $-B(\epsilon_0)$. The form of the hole energy, $\varepsilon^h(\lambda)$ gives us that \cite{CR2}
\begin{eqnarray}\label{B}
B(\epsilon_0)=\log{\left(\alpha\frac{\mathcal{D}}{\epsilon_0}\right)}
\end{eqnarray}
where $\alpha$ is a constant.

  Considering now the dot part of the Bethe equations, the dot contribution to the density satisfies, 
\begin{eqnarray}
f_1(\lambda-2c)&=&\rho^\text{d}_1(\lambda)+\int^\infty_{-B}a_2(\lambda-\mu)\rho^\text{d}_1(\mu),\\
{\rm with~~~~~~} f_n(x)&=&\frac{1}{2\pi}\int_{-\infty}^\infty e^{i\omega x}\frac{\sinh{(\pi-n\phi)\omega}}{\sinh{2\pi\omega}}.
\end{eqnarray}
The solution  is obtained by the Wiener-Hopf method  (see \cite{TWAKM},\cite{RMP} or \cite{Takahashi} and references
therein). Upon integrating over the result as in \eqref{ni} we find that the exact dot occupation in the ground state is,
\begin{eqnarray}\nonumber
n_d&=&\frac{-i}{2\sqrt{\pi}}\int_{-\infty}^\infty  d \omega \frac{e^{-i\omega(2\log{\left(\frac{\epsilon_0}{\Gamma}\right)}+a)}}{ \sinh{(2\pi\omega)}}\\\label{n1}&&
\times\frac{\Gamma(\frac{1}{2}+i(K-1))\omega)}{\Gamma(1+i\omega)\Gamma(1-i(2-K)\omega)}.
\end{eqnarray}
where $\Gamma(x)$ is the Gamma function, $a$ is a non-universal constant and we have used \eqref{K} to write $n_d$ in terms of the Luttinger $K$. As there is no dependence on the cutoff  we can safely take the universal limit $\mathcal{D}\to\infty$ while holding the level width $\Gamma$ fixed. The width serves as both the coupling constant and as  the strong coupling scale  paramerizing the model, with respect to which all quantities are measured. It appears here, rather surprisingly, unrenormalized by the interactions which are present in the system and independent of the raw cut-off, unlike the  case for a dot placed on the boundary \cite{CR2}. We will comment on this further in the next section but for now we examine the expression \eqref{n1}. First we can check that upon inserting $K=1$ in the above expression we recover the non interacting result 
\begin{eqnarray}\label{ndnon}
n_d=\frac{1}{2}-\frac{1}{\pi}\arctan{\left(\frac{\epsilon_0}{\Gamma}\right)}.
\end{eqnarray}
For other values we  may evaluate \eqref{n1} by contour integration and obtain an expansion of $n_d$ for $\epsilon_0<\Gamma$ or $\epsilon_0>\Gamma$ giving
\begin{eqnarray}\label{ng1}
n_d=\begin{cases}
\frac{1}{2}-\left[\sum_{n=0}^\infty a_n\left(\frac{\epsilon_0}{\Gamma}\right)^{2n+1}+b_n\left(\frac{\epsilon_0}{\Gamma}\right)^{(2n+1)/(K-1)}\right]\\
\sum_{n=0}^\infty c_n\left(\frac{\Gamma}{\epsilon_0}\right)^{n+1}~~~~~~~~~~~~~~~~~~~~{\rm for~~}  \Gamma< \epsilon_0
\end{cases}
\end{eqnarray} 
where $a_n,b_n$ and $c_n$ are constants. Furthermore the capacitance of the dot is
\begin{eqnarray}\label{chi}
\chi=\left.\pd{n_d}{\epsilon_0}\right|_{\epsilon_0=0}=\frac{1}{\pi(K-2)\Gamma}.
\end{eqnarray}
 \begin{figure}
\includegraphics[width=.23\textwidth]{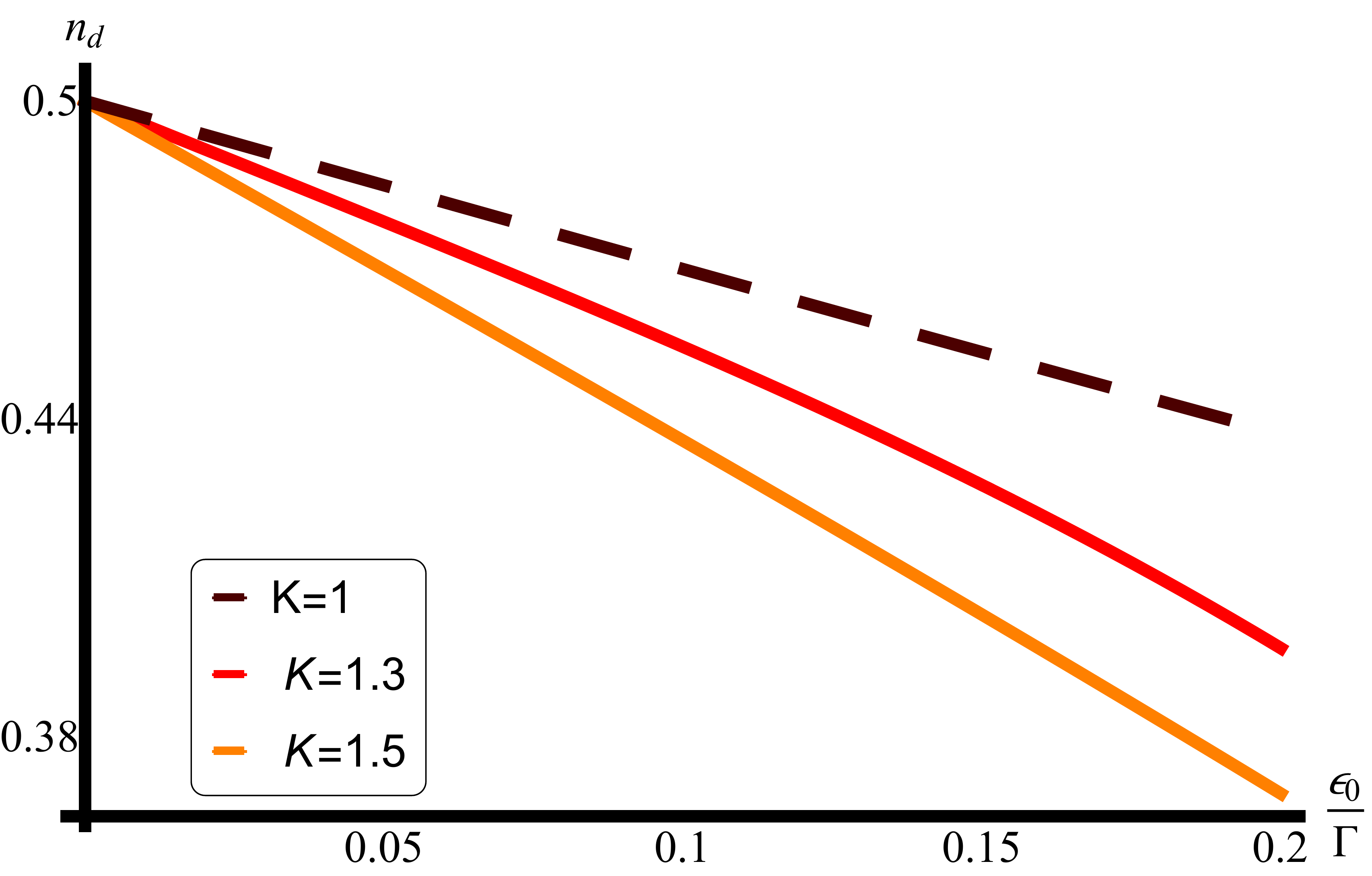}
\includegraphics[width=.23\textwidth]{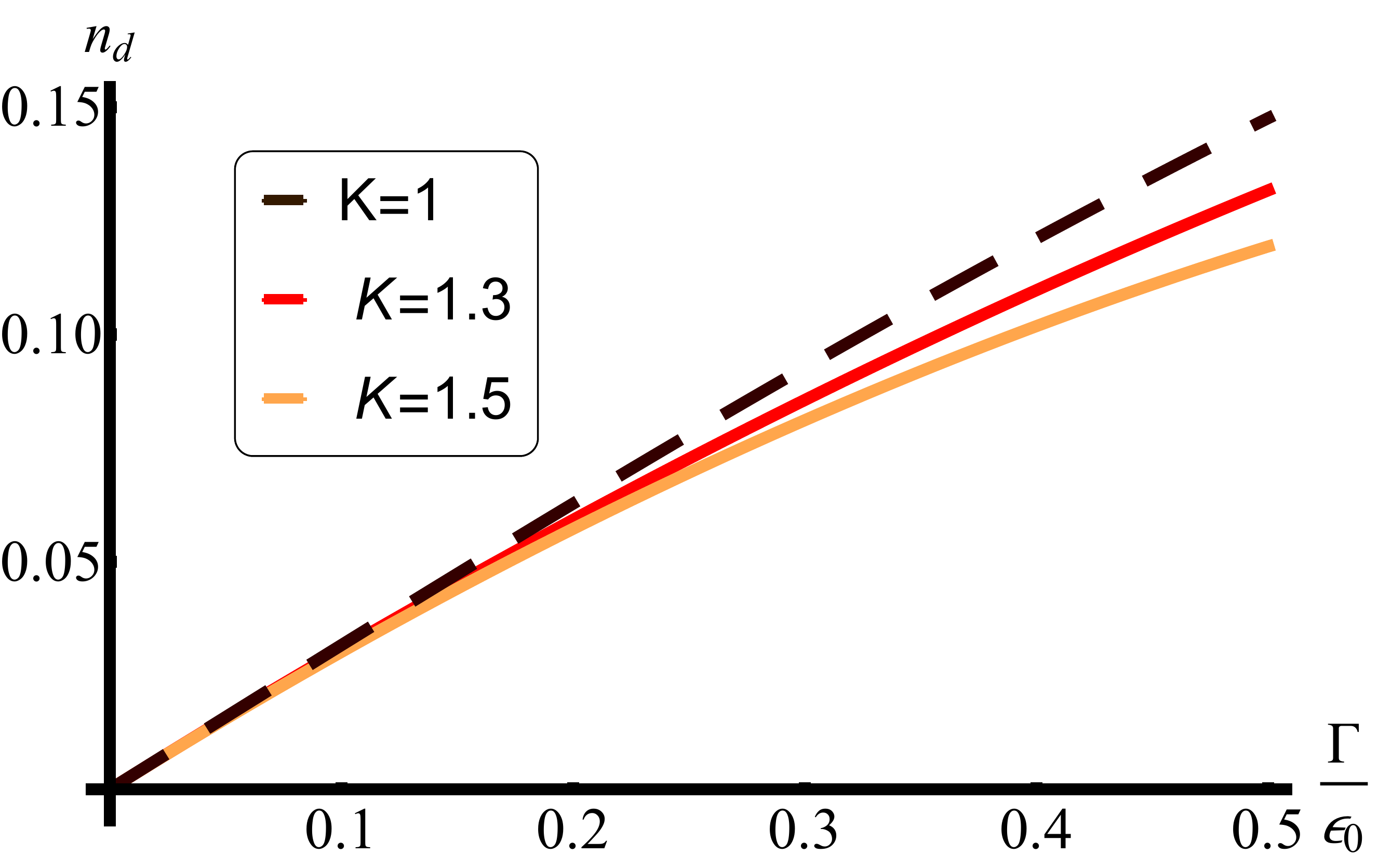}
\caption{(Color Online). The dot occupation at small (left) and large (right) dot energy, $\epsilon_0/\Gamma$, for different values of $K>1$. The effect of attractive interactions is to suppress the dot occupation as compared to the non interacting case (dashed line). This effect becomes stronger for increasing $K$.}
\end{figure} 
We see that at low energy, $\epsilon_0<\Gamma$ the system is strongly coupled with the dot  becoming hybridized with the bulk. At the low energy fixed point $(\epsilon_0=0)$ the dot is fully hybridized and has $n_d=1/2$. The leading term in the expansion about this is $\epsilon_0/\Gamma$ which indicates that the leading irrelevant operator has dimension 2. We identify it as the stress energy tensor \cite{ZAM2}. The next order term $(\epsilon_0/\Gamma)^{1/(K-1)}$ is due to the backscattering which is generated at low energies but is irrelevant for $K>1$. At high energies, $\epsilon_0>\Gamma$, the system becomes weakly coupled with the fixed point ($\epsilon_0\to\infty$) describing a decoupled empty dot, $n_d=0$. The expansion about this fixed point is in terms of integer powers  indicating  that the tunnelling operator $d^\dag \psi(0)$ has dimension $1/2$. The first few terms of the expansion are plotted in Fig. 5 from which we see that the dot occupation is suppressed as a function of $\epsilon_0$ for $K>1$ as compared to the non interacting case due to the backscattering.

\subsection{$K<1$}
The ground state takes a different form in the region $\phi\in[-\pi,0]$ which corresponds to $K\in[0,1]$ . It  is constructed by taking the chiral parameters $\lambda_j\in\mathbb{R}$ to be real  and the rapidities placed on the $2\pi i$ line i.e. Im$(z_\alpha)=2\pi$. Inserting these values into the Bethe equations and then passing to the continuous form we obtain a set of coupled integral equations for the distributions of the charge, $\rho_-(z_j)=1/L(z_j-z_{j-1})$ and chiral variables $\sigma_1(\lambda_j)=1/L(\lambda_j-\lambda_{j-1})$ which we can again split into bulk and dot contributions. The bulk contributions $\rho^\text{b}_-(z)$ and $\sigma_1^\text{b}(\lambda)$ are governed by the continuous Bethe equations,
\begin{eqnarray}\nonumber
\frac{\mathcal{D}e^{z/2}}{4\pi}&=&\rho_-^\text{b}(z)-\int_{-B'}^\infty a_1(z-y)\sigma_1^\text{b}(y)\\
\int_{-B'}^{\infty}a_1(\lambda-y)\rho_-^\text{b}(y)&=&\sigma_1(\lambda)+\int_{-\infty}^\infty a_2(\lambda-y)\sigma_1^\text{b}(y)
\end{eqnarray}
where the  rapidities are bounded by $-B'(\epsilon_0)$. When the dot energy is set to zero we have that $B'(0)=\infty$ and the bulk ground state distributions are found to be,
\begin{eqnarray}
\rho^\text{b}_-(z)&=&\frac{\mathcal{D}e^{z/2}}{2\pi},\\
\sigma^\text{b}_1(\lambda)&=&\frac{\mathcal{D}e^{z/2}}{4\pi\cos{(\phi/2)}}.
\end{eqnarray}
The fundamental excitations above this ground state consist of adding holes to either of these distributions. The energy of these are $\varepsilon^h(z)=4\pi\rho^\text{b}_-(z)$ and $\varepsilon^h(\lambda)=4\pi\sigma^\text{b}_1(\lambda)$ for a charge hole and chiral hole respectively. As in the previous section these are used to determine $B'$ which gives the same relation as \eqref{B}.  
\begin{figure}
\includegraphics[width=.23\textwidth]{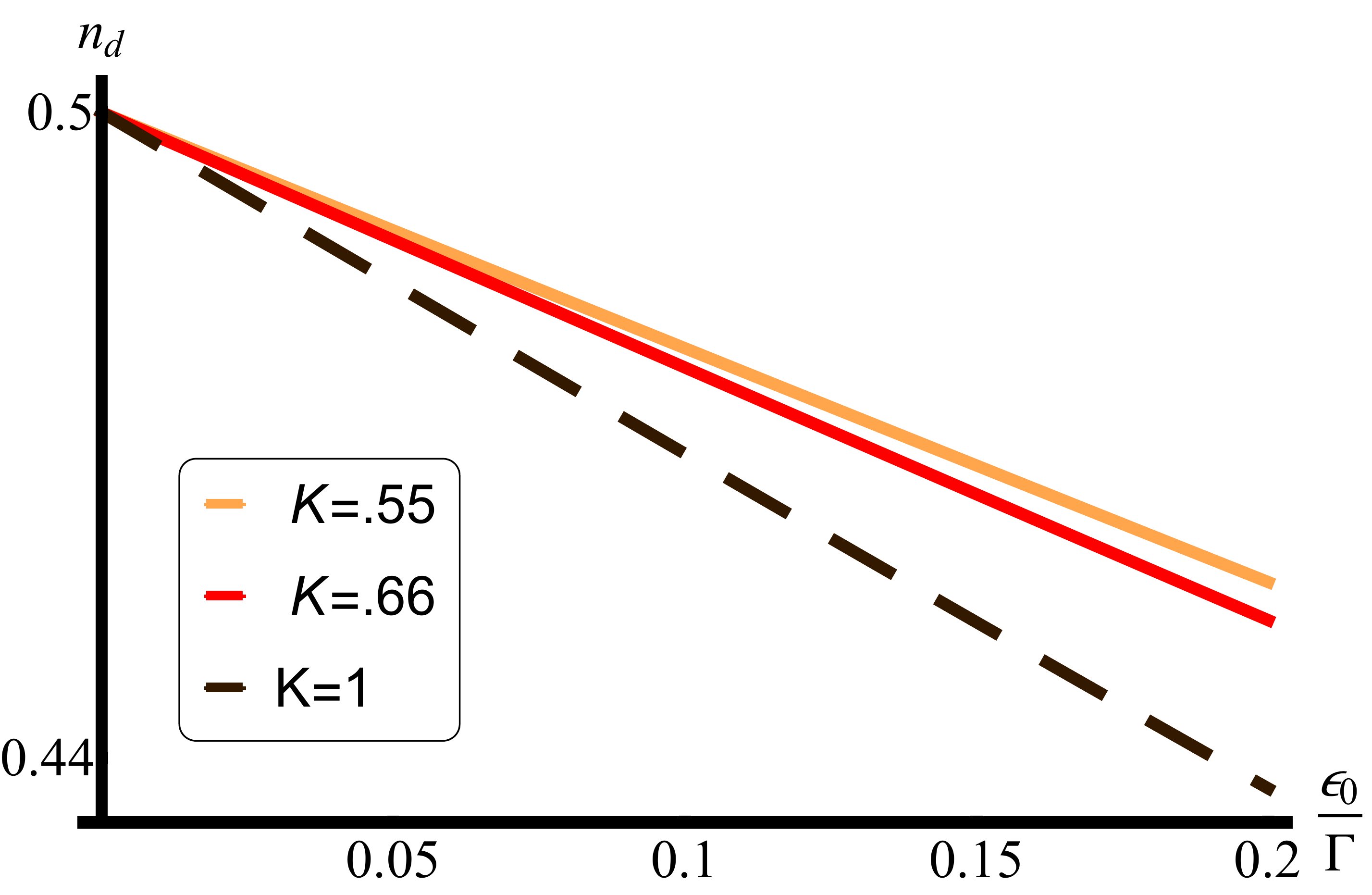}
\includegraphics[width=.23\textwidth]{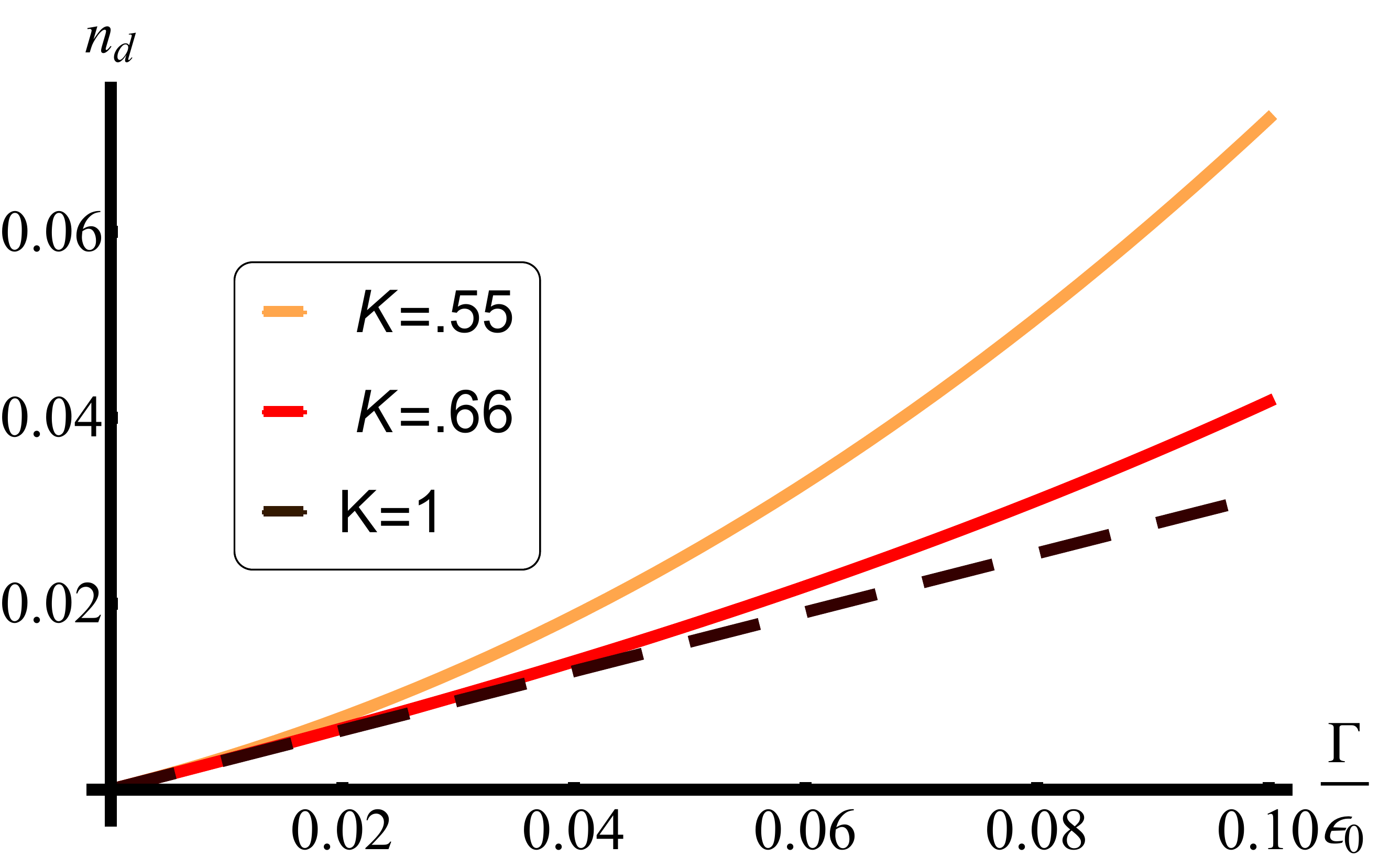}
\caption{(Color Online).The dot occupation at small (left) and large (right) dot energy for different values of $K$. The effect of repulsive interactions $K<1$ is to enhance the dot occupation as compared to the non interacting case (dashed line) with the effect increasing as $K$ decreases.}
\end{figure} 
 The dot occupation is subsequently obtained by integrating over the dot part of the charge distribution $n_d=\int\rho^\text{d}_-(z)\mathrm{d}z$ which is determined by,
\begin{eqnarray}
g_2(\lambda-2c)&=&\rho^\text{d}_-(\lambda)+\int^\infty_{-B'}g_1(\lambda-y)\rho^\text{d}_-(y),~~~~~\\
g_n(x)&=&\frac{1}{2\pi}\int_{-\infty}^\infty e^{i\omega x}\frac{\sinh{(\pi-\phi)\omega}}{2\cosh{(\phi\omega)}\sinh{(n\pi\omega)}}.~~
\end{eqnarray}
The solution is again determined using the Wiener-Hopf method with the result that the dot occupation for $K<1$ is also given by  \eqref{n1}. Note however that the poles at $\omega=i(K-1)(2n+1)/2$ have shifted from the upper half plane to the lower half plane. This changes the expansions at high and low energy to be 
\begin{eqnarray}\label{nl2}
n_d=\begin{cases}
\frac{1}{2}-\sum_{n=0}^\infty a_n\left(\frac{\epsilon_0}{\Gamma}\right)^{2n+1}\\
\sum_{n=0}^\infty c_n\left(\frac{\Gamma}{\epsilon_0}\right)^{n+1}+b_n\left(\frac{\Gamma}{\epsilon_0}\right)^{(2n+1)/(1-K)}
\end{cases}
\end{eqnarray} 
with the capacitance being given by \eqref{chi}. As in the $K>1$ region, the dot is strongly coupled at low energy and weakly coupled at high energy with the same leading terms in the expansion about these points however the term generated by the backscattering now appears in the expansion about the high energy fixed point. This stems from the fact that backscattering is relevant for $K<1$ and leads to an enhancement of the dot occupation as compared to the $K=1$ case, see Fig. 6. 

The dot occupation for the embedded system is simply obtained from \eqref{n1} by using the mapping $K\to 1/K$.
\begin{figure}
\includegraphics[trim = 20mm 20mm 20mm 10mm, clip, width=.46\textwidth]{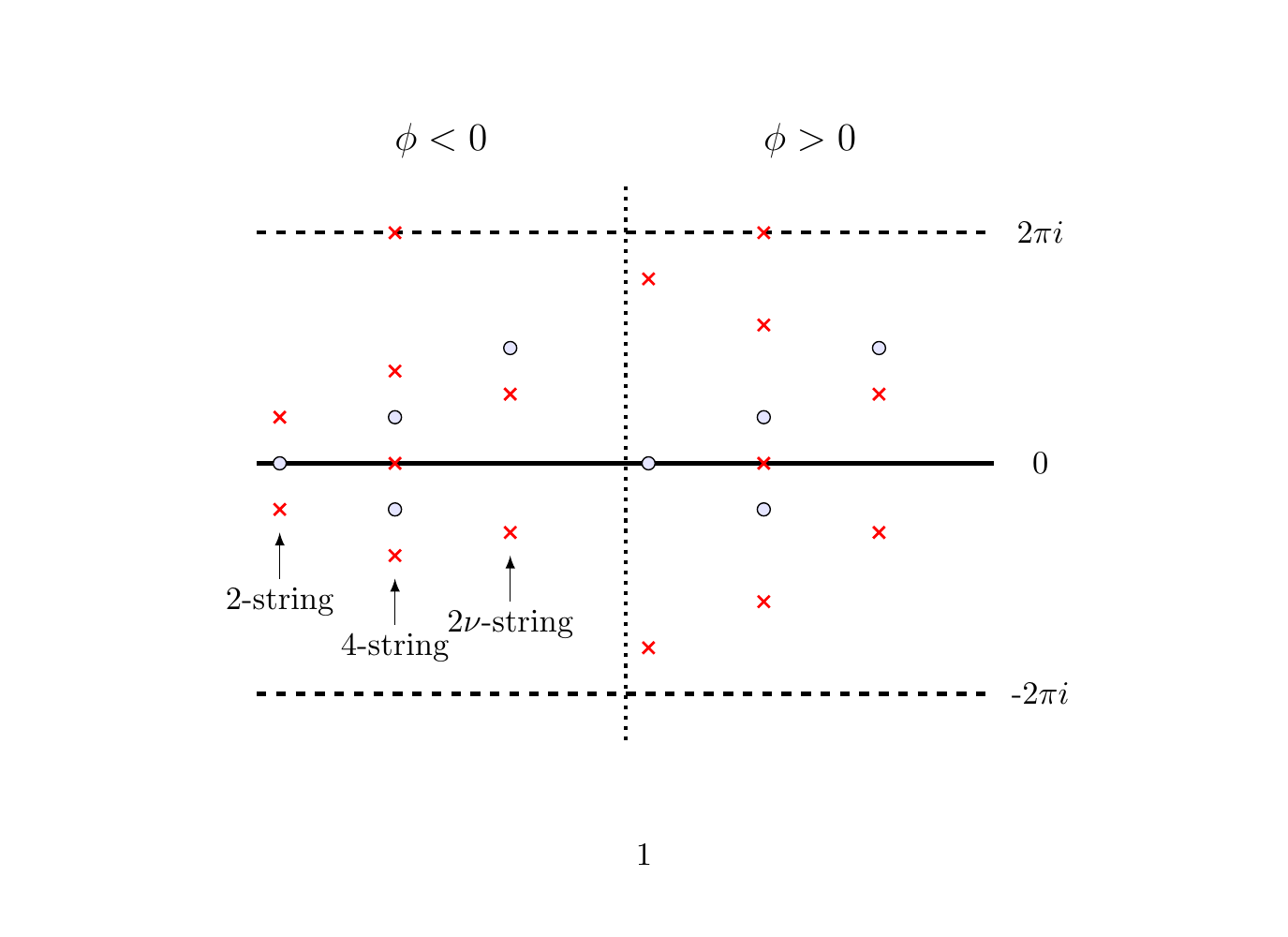}
\caption{(Color Online) At finite temperature the rapidity and chiral variables may form $z-\lambda$ strings where $n$ $\lambda$s and $2n$ $z$s form a set given by \eqref{zlam}. On the left we show how a $2$-string, $4$-string and the negative parity $2\nu$-string are arranged for $\phi<0$. On the right we depict the same for $\phi>0$. Note only the $z$ positions are changed when going from left to right which results in a change in sign of the energy from the strings.}
\end{figure} 
\section{RG flow}
In the previous section we derived exact expressions for the dot occupation for the side-coupled model as a function of $\epsilon_0$ measured with respect to the strong coupling scale. This strong coupling scale is given by $\Gamma$, the level width. It does not depend on $K$ as might be expected for an interacting model and in fact coincides with the free model. 
To understand why the level width is not renormalised by $K$ we can make use of the mapping to the embedded model. The strong coupling scale in the embedded model should behave similarly to the single lead case, where a dot is placed at a Luttinger liquid edge \cite{Duality}. For an arbitrary Coulomb interaction, $U$ this is $\mathcal{D}(\Gamma/\mathcal{D})^{1/\alpha}$ where $\alpha=1+2\left[\arctan{(g)}-\arctan{(U)}\right]/\pi$ \cite{CR2}. Taking $U=g$, as required by the mapping (see \eqref{Hc}), reduces this to $\Gamma$, the free value. The non-renormalization of the level width suggests that the tunnelling operator $d^\dag\psi_\pm(0)$ should have the same dimension as the free model which is confirmed by the high energy expansions of the dot occupation. This is in stark contrast to the the fact that fermions in a Luttinger liquid (away from the edge) have dimension $(K+1/K)/4$. Thus the remarkably simple expression for the strong coupling scale and critical exponents present here stand in contrast to a quite substantial modification of the fermions in the vicinity of the dot.

We now have the following picture of the side-coupled system. For all $K\in[0,2]$ the system flows from weak coupling at high energy to strong coupling at low energy. The low energy fixed point describes a dot which is fully hybridized with the bulk and has the fixed point occupation $n_d=1/2$. The hybridized dot then acts as a backscattering potential via co-tunnelling. The leading irrelevant operator which perturbs away from the fixed point is the stress energy tensor and results in odd integer powers of $\epsilon_0/\Gamma$ in the dot occupation. For $K>1$ the backscattering is irrelevant which gives rise to odd powers of  $(\epsilon_0/\Gamma)^{1/(K-1)}$ resulting in a suppression of the dot occupation at $\epsilon_0>0$.  For $K<1$ on the other hand it is relevant and generates no other terms in the expansion. The high energy fixed point describes a decoupled dot which has $n_d=0$ for $\epsilon_0\to \infty$ or $n_d=1$ for $\epsilon_0\to-\infty$. By reducing the energy scale we flow away from the fixed point with the tunnelling operator $d^\dag\psi_\pm(0)$ which is the leading relevant operator and has dimension $1/2$ as in the free model. This give rise to integer powers of $\Gamma/\epsilon_0$ in $n_d$. Additionally when $K<1$ backscattering is relevant and causes  odd powers of $(\Gamma/\epsilon_0)^{1/(1-K)}$ to appear resulting in an enhancement of the dot occupation .

\section{Thermodynamics}
\subsection{$K=\frac{\nu-1}{\nu}$}
In this section we study the finite temperature properties of the dot by calculating the free energy. To do so we use the  methods developed by Yang and Yang \cite{YY} and later extended by Takahashi \cite{Takahashi} based on the string hypothesis. This states that in the thermodynamic limit the solutions of the Bethe equations take complex values organised into strings. The form of the strings depend upon the model and the values of the parameters therein. To simplify matters we take $\phi=\pm\pi/\nu$ with $\nu$ an integer so that $K=\frac{\nu\pm 1}{\nu}$. With this value fixed the hypothesis states that the Bethe equations allow for the following forms of the charge and chiral variables. 
\begin{figure}
\includegraphics[width=.46\textwidth]{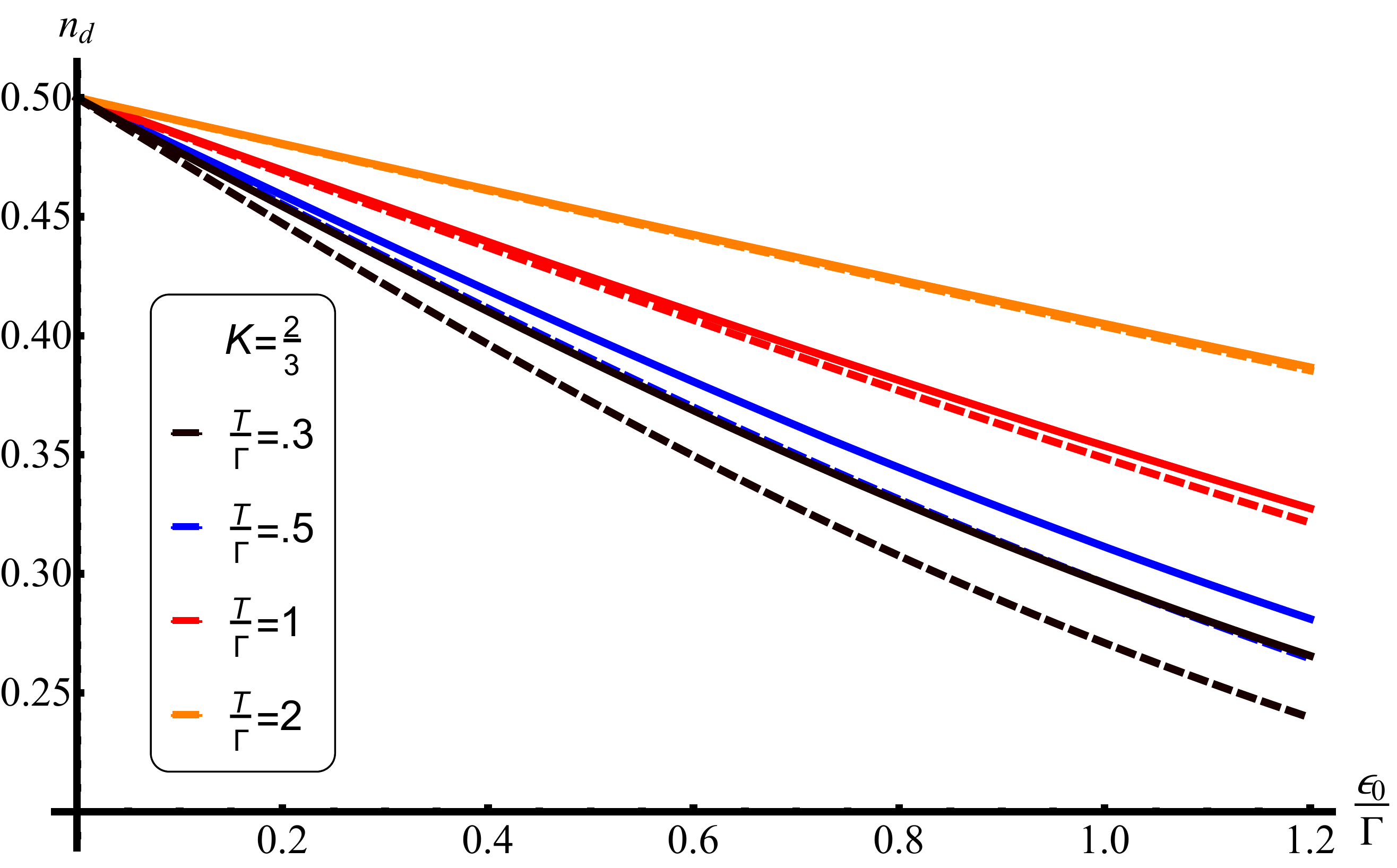}
\includegraphics[width=.46\textwidth]{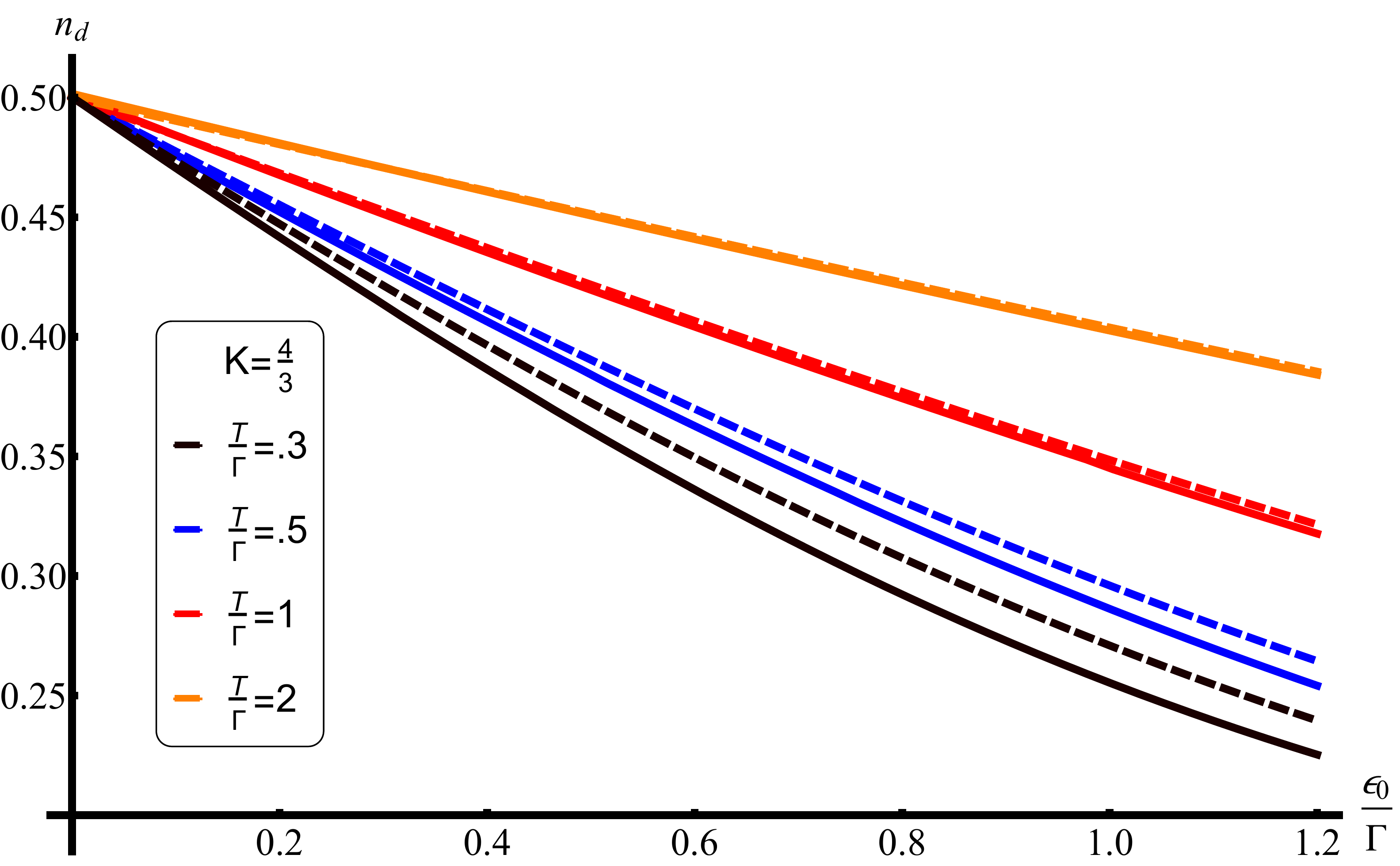}
\caption{(Color Online): The finite temperature dot occupation is plotted as a function of $\epsilon_0/\Gamma$ for several values of the temperature. Above we plot the dot occupation with $K=\frac{2}{3}$ (solid lines) and $K=1$ (dashed lines). The repulsive bulk interactions result in an enhancement of the dot occupation in comparison to the non interacting case. This is effect is most pronounced for lower temperatures. At higher temperature the interacting and non interacting curves coincide owing to the fact that the dot becomes decoupled. Below we plot the same for $K=\frac{4}{3}$ (solid lines) and plot again $K=1$ (dashed) for comparison. The dot occupation is suppressed  due to the attractive interactions wth the effect becoming more pronounced for lower $T/\Gamma$.}
\end{figure} 
\begin{figure}
\includegraphics[width=.46\textwidth]{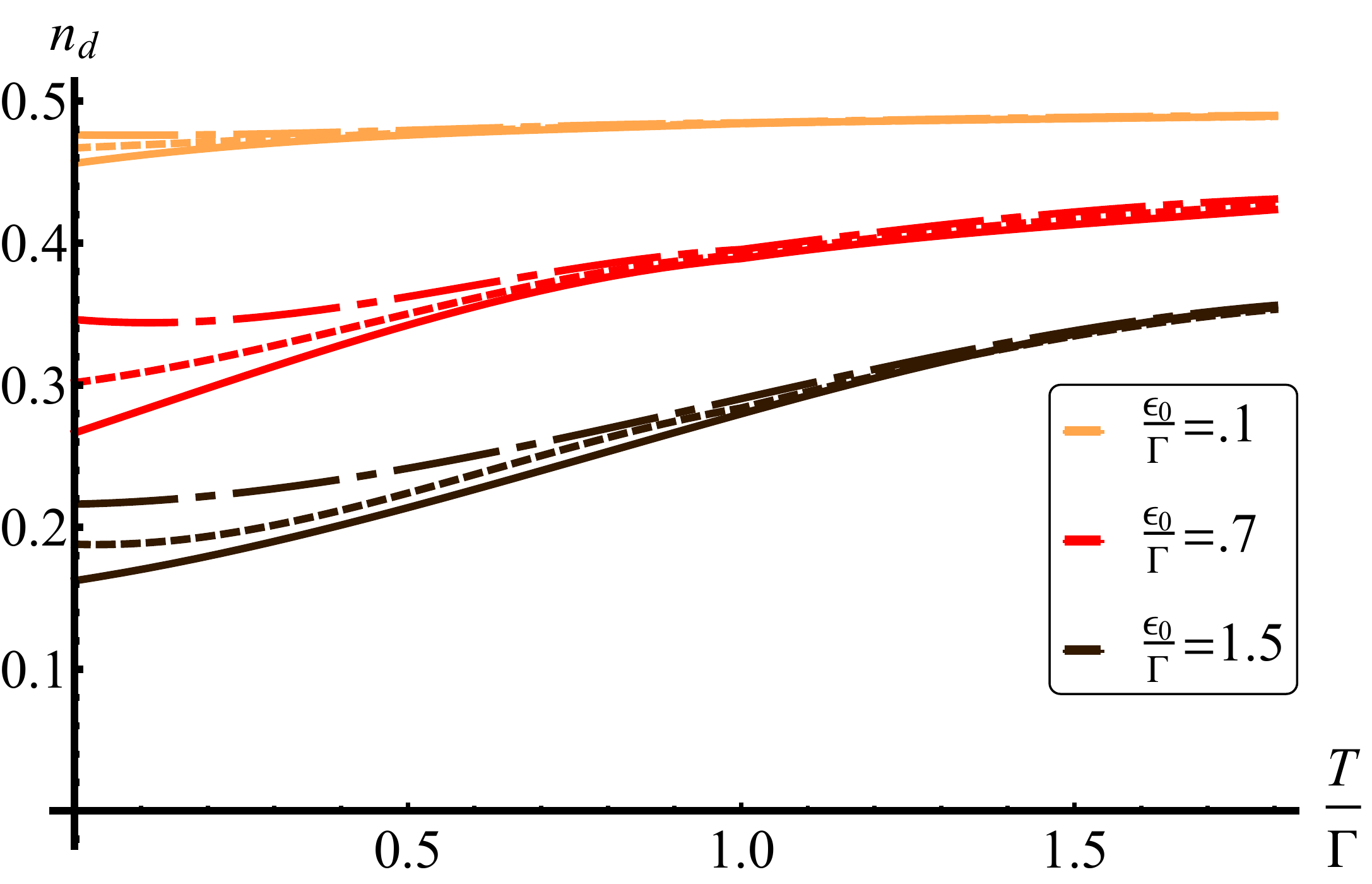}
\caption{(Color Online): The dot occupation for fixed $\epsilon_o/\Gamma$ as a function of temperature. The interaction is taken to be $K=\frac{4}{3}$ (dot-dashed lines), $K=1$ (dashed lines) and $K=\frac{2}{3}$ (solid lines). We see the enhancement and suppression of the dot occupation for repulsive and attractive interaction with the effect most pronounced as the temperature is lowered.}
\end{figure}
 
The rapidities can be real or complex with Im$(z)=0,2\pi$. These contribute bare energy $\pm\mathcal{D}e^{z/2}$ and we denote the distributions of these $\rho_\pm(z)$. The chiral variables can take on complex values so that they arrange into $n$-strings with $n<\nu$ such that 
\begin{eqnarray}
\lambda^{(n)}_l=\lambda^{(n)}+i\phi(n-1-2l),~~~l=0,\dots,n-1
\end{eqnarray}
or $\lambda$ on the $i\pi$ line which is sometimes called a negative parity string. The $\lambda$ $n$-strings have no bare energy and we denote the distributions of their real part, called the string centre $\sigma_n(\lambda)$ with $n=\nu$ denoting the negative parity string. Also possible are $z-\lambda$ $2n$-strings consisting  of $2n$ $z$s and a $\lambda$ $n$-string taking the values
\begin{eqnarray}\label{zlam}
z_{l+1}^{(n)}=\lambda^{(n)}+i\phi(n-2j)+i\pi+\text{sgn}(\phi)i\pi\\
z_{l+n+1}^{(n)}=\lambda^{(n)}+i\phi(n-2l)+i\pi-\text{sgn}(\phi)i\pi
\end{eqnarray}
where $j=0,\dots,n$ and $l=1,\dots,n-1$. These contribute bare energy  $E_n=-2\text{sgn}(\phi)\cos{(n\phi/2)}\mathcal{D}e^{\lambda^{(n)}/2}$. In addition there is also a negative parity $z-\lambda$ string  
\begin{equation}\label{negstr}
\lambda=\lambda^{(\nu)}+i\pi,~~~
z_{1,2}=\lambda^{(\nu)}\pm i(\pi-\phi)
\end{equation} 
which has energy $2\sin{(\phi/2)}\mathcal{D}e^{\lambda^{(\nu)}/2}$. We denote the distributions of the centres of the $z-\lambda$ $2n$-strings by $\rho_n(z)$ with $n=\nu$ indicating the negative parity string.Several string type are depicted in Fig. 7 for both $\phi>0$ and $\phi<0$.
 
Having elucidated the string structure of the model, the free energy is found, as in other Bethe ansatz models following the procedure laid out in \cite{Takahashi}.  The approach
is well known and we just provide the main steps. The free
energy $F=E-TS$, where $E$ is the energy of an arbitrary configuration of strings and $S$
is its associated Yang-Yang entropy, is minimized
with respect to $\rho_\pm,\rho_n$ and $\sigma_n$ which are solutions of the Bethe Ansatz equations. The result
of this minimization gives the thermodynamic Bethe ansatz (TBA) equations which determine the minimum of $F$. Owing to the different string structures for $K$ greater than or less than 1 we consider each region separately.

We start with  $\phi=-\pi/\nu$, corresponding to $K=\frac{\nu-1}{\nu}<1$, describing repulsive interactions. In this region we find the dot contribution to the free energy is  
\begin{eqnarray}\nonumber
F_d=E^0_d-T\int f_0(x+2\log{\left(\frac{T}{\Gamma}\right)})\log{(1+e^{\varphi_-(x)})}\\\nonumber
-T\int f_0*s(x+2\log{\left(\frac{T}{\Gamma}\right)})\log{(1+e^{\varkappa_1(x)})}\\\label{FE}
-T\int s(x+2\log{\left(\frac{T}{\Gamma}\right)})\log{(1+e^{\varkappa_{\nu-1}(x)})}
\end{eqnarray}
where $E^0_\text{d}$ is the ground state energy due to the dot, $s(x)=\text{sech}{(\pi x/2\phi)}/4\phi$ and $*$ denotes the convolution $f*g=\int f(x-y)g(y)\mathrm{d}y$. The thermodynamic functions $\varphi_\pm,\varphi_n$ and $ \varkappa_n$  are related to the distributions $\rho_\pm, \rho_n$ and $\sigma_n$ respectively and are solutions of the TBA equations   which in this case are
\begin{widetext}
\begin{eqnarray}
\varphi_+&=&s*\log{\left(\frac{1+e^{\varphi_1}}{1+e^{\varkappa_{1}}}\right)},~~
\varphi_-=-2e^{x/2}+s*\log{\left(\frac{1+e^{\varphi_1}}{1+e^{\varkappa_{1}}}\right)}\\
\varphi_n&=&s*\log{(1+e^{\varphi_{n-1}})(1+e^{\varphi_{n+1}})(1+e^{-\varphi_{\nu}})^{\delta_{n,\nu-2}}}+\delta_{n,1}s*\log{\left(\frac{1+e^{\varphi_{+}}}{1+e^{\varphi_{-}}}\right)}\\
\varkappa_{n}&=&s*\log{(1+e^{\varkappa_{n-1}})(1+e^{\varkappa_{n+1}})^{1+\delta_{n,\nu-2}}}-\delta_{n,1}\left[\frac{e^{x/2}}{\cos{(\phi/2)}}-s*\log{\left(\frac{1+e^{\varphi_+}}{1+e^{\varphi_-}}\right)}\right]
\end{eqnarray} 
\end{widetext}
along with $\varphi_{\nu-1}=s*\log{(1+e^{\varphi_{\nu-2}})}+\frac{\nu\epsilon_0}{T}=-\varphi_{\nu}+\frac{2\nu\epsilon_0}{T}$ and $
\varkappa_{\nu-1}=s*\log{(1+e^{\varkappa_{\nu-2}})}=-\varkappa_{\nu}$.
Just as in the calculation of the dot occupation in the ground state the above equations are independent of the cutoff which has been removed while holding $\Gamma$ fixed. These expressions give the exact dot free energy of the system in all temperature regimes. Their complicated nature precludes any analytic solution for the thermodynamic functions but are straightforwardly determined numerically through iteration of the integral equations.

Before doing this however we can examine them in the limits of low and high temperature. The functions $f_0(x)$ and $s(x)$ appearing in the free energy are sharply peaked about zero meaning that for $T\to 0,\infty$ the free energy is determined by the solutions of the TBA in the $x\to\infty,-\infty$ limits respectively. Setting $\epsilon_0=0$ and taking first the high temperature limit, $x\to-\infty$ we see that the driving terms in the TBA vanish and the thermodynamic functions are constants  $e^{\varphi_\pm(-\infty)}=1$,
 \begin{eqnarray}\label{TBAh}
 e^{\varphi_j(-\infty)}=e^{\varkappa_j(-\infty)}=(j+1)^2-1\\
 e^{\varphi_{\nu-1}(-\infty)}=e^{\varkappa_{\nu-1}(-\infty)}=\nu-1.
\end{eqnarray}  
Likewise in the opposite low temperature limit $x\to\infty$ we get $e^{\varphi_-(\infty)}=0,~e^{\varphi_+(\infty)}=3$,
 \begin{eqnarray}\label{TBALl}
 e^{\varkappa_j(\infty)}=j^2-1,~ e^{\varkappa_{\nu-1}(\infty)}=\nu-2\\
e^{\varphi_{j}(\infty)}=(j+2)^2-1,~e^{\varphi_{\nu-1}(\infty)}=\nu.
\end{eqnarray}  
The free energy thus becomes linear in $T$ in both the high and low temperature limit.

Using these we can  check the RG picture we arrived at earlier using the ground state dot occupation still holds true at finite temperature. Firstly note that the energy scale, the temperature in this case, is measured with respect to $\Gamma$ which serves as both the strong coupling scale and the level width for the model. Thus the system is strongly coupled at  low temperature $T\ll\Gamma$ and weakly coupled at high temperature $T\gg\Gamma$. Furthermore by inserting \eqref{TBALl} \eqref{TBAh} into \eqref{FE} we obtain the  $g$-function of the model, defined to be the difference in the UV and IR entropy of the impurity
\begin{eqnarray}
g=S_{\text{UV}}-S_{\text{IR}}=\log{2}+\frac{1}{2}\log{\left(\frac{1}{K}\right)}.
\end{eqnarray}
This is always positive for the range of values considered in agreement with the requirement that as we move along the RG flow by lowering the temperature,  massless degrees of freedom are integrated out. The first term comes from the charge degrees of freedom and corresponds to the entropy of a decoupled dot at high temperature which is fully hybridised at low temperature. The second term comes from the chiral degrees of freedom and is the same as for the Kane-Fisher model of a back scattering impurity\cite{FSW}\cite{ryl}. We see from this that at high temperature the dot is decoupled and as $T$ is lowered it becomes hybridised with the dot whereupon it acts as a back scattering impurity. In the non interacting limit the $K\to1$ this last term disappears and we recover the expected result. 

We may go beyond the fixed point behaviour to get the leading order corrections and determine the specific heat. Following \cite{TWAKM}\cite{deSa} we expand about the low temperature solution $\log{(1+\exp{\varphi_-})}\approx \exp{-2e^{x/2}}$ and $\log{(1+\exp{\varkappa_1})}\approx \exp{-e^{x/2}/\cos{(\phi/2)}}$ for $x\gg 0$. The low temperature specific heat is then found to be
\begin{eqnarray}
C_v\sim \frac{T}{\Gamma}
\end{eqnarray}
 which agrees with the expectation that the irrelevant operator is the stress energy tensor.

By numerically integrating the TBA and using them in \eqref{FE} we can obtain the finite temperature dot occupation of the system. This is plotted in Fig. 8 for $K=\frac{2}{3}$ as a function of $\epsilon_0/\Gamma$ at different values of the temperature, $T/\Gamma$. For the same value of $K$ we plot the dot occupation at fixed $\epsilon_0/\Gamma$ as a function $T/\Gamma$ in the Fig. 9. Comparing to the dashed lines which are the non interacting values we see that the dot occupation is enhanced just as it was at zero $T$. This enhancement is strongest at low $T$ and is washed out at high temperature as the system becomes weakly coupled.

\subsection{$K=\frac{\nu+1}{\nu}$}

We turn now to the case of $\phi=\pi/\nu$ or $K=\frac{\nu+1}{\nu}>1$, attractive interactions. In this regime we will see that tunnelling to the dot is still relevant however it must compete with the backscattering that this generates which is irrelevant for $K>1$\cite{KF}. This competition makes itself felt via changes in the free energy and TBA equations. The dot contribution to the free energy is now given by
\begin{eqnarray}\nonumber
F_d=E^0_d-T\int f_0(x+2\log{\left(\frac{T}{\Gamma}\right)})\log{(1+e^{-\varphi_+(x)})}\\\nonumber
-T\int f_0*s(x+2\log{\left(\frac{T}{\Gamma}\right)})\log{(1+e^{\varphi_1(x)})}\\
-T\int s(x+2\log{\left(\frac{T}{\Gamma}\right)})\log{(1+e^{\varkappa_{\nu-1}(x)})}
\end{eqnarray}
with the TBA equations being
\begin{widetext}
\begin{eqnarray}
\varphi_+&=&2e^{x/2}+s*\log{\left(\frac{1+e^{\varphi_1}}{1+e^{\varkappa_{1}}}\right)},~~
\varphi_-=s*\log{\left(\frac{1+e^{\varphi_1}}{1+e^{\varkappa_{1}}}\right)}\\
\varphi_n&=&s*\log{(1+e^{\varphi_{n-1}})(1+e^{\varphi_{n+1}})(1+e^{-\varphi_{\nu}})^{\delta_{n,\nu-2}}}-\delta_{n,1}\left[s*\log{\left(\frac{1+e^{-\varphi_{+}}}{1+e^{-\varphi_{-}}}\right)}+\frac{e^{x/2}}{\cos{(\phi/2)}}\right]\\
\varkappa_{n}&=&s*\log{(1+e^{\varkappa_{n-1}})(1+e^{\varkappa_{n+1}})^{1+\delta_{n,\nu-2}}}-\delta_{n,1}s*\log{\left(\frac{1+e^{-\varphi_+}}{1+e^{-\varphi_-}}\right)}
\end{eqnarray} 
\end{widetext}
and $\varphi_{\nu-1}=s*\log{(1+e^{\varphi_{\nu-2}})}+\frac{\nu\epsilon_0}{T}=-\varphi_{\nu}+\frac{2\nu\epsilon_0}{T}$ as well as $
\varkappa_{\nu-1}=s*\log{(1+e^{\varkappa_{\nu-2}})}=-\varkappa_{\nu}$. Comparing to the $K<1$ case we see that the roles of $e^{\phi_-}$ and $e^{-\phi_+}$ have been exchanged and that the exponential driving term now appears in the $\varphi_{j}$ equations rather than $\varkappa_j$ ones. 

We gain insight to the $K>1$ region by looking at the asymptotic solutions of the TBA. The high temperature solutions, $x\to-\infty$ remain unchanged and are given by \eqref{TBAh}, therefore as $T\to\infty$ the system is the same regardless of $K$. In the low temperature limit however the solutions are different as should be the case given the ground state is of a different form. We get that $e^{-\varphi_+(\infty)}=0,~e^{\varphi_-(\infty)}=3$,
 \begin{eqnarray}
 e^{\varphi_j(\infty)}=j^2-1,~ e^{\varphi_{\nu-1}(\infty)}=\nu-2\\
e^{\varkappa_{j}(\infty)}=(j+2)^2-1,~e^{\varkappa_{\nu-1}(\infty)}=\nu
\end{eqnarray}  
Using these in the $g$ function we obtain the same form as before,
\begin{eqnarray}
g=\log{2}+\frac{1}{2}\log{\left(\frac{1}{K}\right)}.
\end{eqnarray}
Note however that although $g>0$,  the second term which is due  to the backscattering,  is negative for $K>1$. This relative sign between the charge and chiral terms is related to the competition between the tunnelling and the backscattering. Upon taking the $K\to1$ we recover the non interacting result. The low temperature corrections to the fixed point can be obtained as they were in the previous section. This time however the driving terms in the do not appear in the $\varkappa_1$ equation but in the $\varphi_1$ equation instead and consequently  we take $\log{(1+\exp{-\varphi_+})}\approx \exp{-2e^{x/2}}$ and $\log{(1+\exp{\varphi_1})}\approx \exp{-e^{x/2}/\cos{(\phi/2)}}$ for $x\gg 0$ and find the specific heat to be 
\begin{eqnarray}
C_v\sim \frac{T}{\Gamma} +a \left(\frac{T}{\Gamma
}\right)^{\alpha}`.
\end{eqnarray}
Again the leading order term coincides with the stress tensor being the leading irrelevant operator. The term scales as $T^\alpha$ where $\alpha=2$ for $K=\frac{\nu+1}{\nu},~\nu>2$. It is expected however that $\alpha$ becomes non integer when increasing $K$ beyond this as is the case in the ground state dot occupation. 

The finite temperature dot occupation can be obtained by numerically integrating the TBA as in the previous section and the results are plotted in Fig. 8 and Fig. 9. We see that the dot occupation is suppressed as compared to $K=1$ or $K<1$, with the effect being most pronounced at low temperature. At high $T$ the dot becomes decoupled and the occupation approaches that of the non interacting case.

\section{Conclusion}
In this article we have solved two related models of quantum dots coupled to Luttinger liquids. The first consists of a dot side-coupled to the Luttinger liquid while in the second the dot is placed between two otherwise disconnected liquids. The latter also requires that a Coulomb interaction between the occupied dot and the end of the liquids is included and it is tuned to the same value as the bulk interaction. The side-coupled  model however, requires no such tuning. 

The solution shows that the two models are related by taking $K\to 1/K$ which was shown previously through bosonization \cite{Duality}. We derived the Bethe equations for both models and used them to construct the ground state and derive exact expressions for the dot occupation in all parameter regimes. It was seen that the side-coupled system is strongly coupled at low energies so that the dot becomes fully hybridised with the bulk and acts as a backscattering potential. The effect of the backscattering is to either suppress or enhance the dot occupation depending on the sign of the interactions. 

The scaling dimensions of the leading relevant and irrelevant operators about the UV and IR fixed points were found to coincide with that of the free model. The surprising result that the fixed points appear, at least to leading order to be Fermi liquid is in start contrast to the non-Fermi liquid nature of the bulk system.

We then examined the finite temperature properties of the dot by deriving the Thermodynamic Bethe equations and free energy of the system. It was seen that at low temperature dot is fully hybridised with the bulk and the interactions resulting in a suppression or enhancement of the dot occupation. The effect of the interactions is washed out at high temperature whereupon the dot decouples.

The lack of fine tuned parameters in the side-coupled model make it a good candidate for experimental realizations. Such a system may be created placing a quantum dot near a carbon nanotube,  the edge of a quantum Hall sample or a topological insulator. The dot occupation can then be measured by means of a quantum point contact and compared to \eqref{n1}.

\acknowledgements{ We are grateful to Yashar Komijani and Moshe Goldstein for useful discussions and comments. CR is supported by the Peter Lindenfeld Fellowship and NA by NSF Grant DMR 1410583}.
\bibliography{mybib}
\end{document}